\documentclass[aps,floatfix,prd,twocolumn]{revtex4}
\usepackage{graphicx}
\usepackage{dcolumn}
\usepackage{bm}
\usepackage{float}
\usepackage{placeins}
\usepackage{amsmath}
\usepackage{romannum}
\usepackage{csquotes}
\usepackage{hyperref}
\hypersetup{
    colorlinks = true,
    citecolor = blue,
    linkcolor = blue,
    urlcolor = blue,
}

\begin{document}

\title{Non-equilibrium thermoelectric transport across normal metal-Quantum dot-Superconductor hybrid system within the Coulomb blockade regime}
\author{Sachin Verma}
\author{Ajay Singh}
\affiliation{Department of Physics, Indian Institute of Technology, Roorkee, Uttarakhand, 247667, India\\
\text{\bf{Email:}} sverma2@ph.iitr.ac.in and ajay@ph.iitr.ac.in}

\begin{abstract}
A detailed investigation of the non-equilibrium steady-state electric and thermoelectric transport properties of a quantum dot coupled to the normal metallic and s-wave superconducting reservoirs (N-QD-S) are provided within the Coulomb blockade regime. Using non-equilibrium Keldysh Green's function formalism, initially, various model parameter dependences of thermoelectric transport properties are analysed within the linear response regime. It is observed that the single-particle tunnelling close to the superconducting gap edge can generate a relatively large thermopower and figure of merit. Moreover, the Andreev tunnelling plays a significant role in the suppression of thermopower and figure of merit within the gap region. Further, within the non-linear regime, we discuss two different situations, i.e., the finite voltage biasing between isothermal reservoirs and the finite thermal gradient in the context of thermoelectric heat engine. In the former case, it is shown that the sub-gap Andreev heat current can become finite beyond the linear response regime and play a vital role in asymmetric heat dissipation and thermal rectification effect for low voltage biasing. The rectification of heat current is enhanced for strong on-dot Coulomb interaction and at low background thermal energy. In the latter case, we study the variation of thermovoltage, thermopower, maximum power output, and corresponding efficiency with the applied thermal gradient. These results illustrate that hybrid superconductor-quantum dot nanostructures are promising candidates for the low-temperature thermal applications.\\
\\
\textbf{\textit{KEYWORDS --- quantum dot, superconductivity, Andreev bound states, Coulomb blockade, Keldysh formalism, linear and non-linear thermoelectric transport, Seebeck effect, thermoelectric heat engine, Peltier effect, thermal diode}}
\end{abstract}
\maketitle
\pagenumbering{arabic}

\section{Introduction}
Thermoelectric materials convert heat into electricity and electricity into the temperature difference based on the Seebeck and Peltier effects, respectively. These abilities of thermoelectric materials could have many applications, including power generation and solid-state refrigeration \cite{Tritt2006,Bell2008,Enescu2019,Jaziri2020,Zoui2020}. Recently, there has been a growing interest in identifying and utilizing the materials or systems that convert heat energy into productive applications as efficiently as possible (i.e., a large output power at the cost of less input heat energy). Systems with a large thermoelectric efficiency ($\eta$) or thermoelectric figure of merit ($ZT$) could be used to develop practical thermoelectric heat engines, or power generators \cite{Mahan1997,Wei2020}. To achieve a high $\eta$ or $ZT$, one requires a high thermopower $S$, a high electrical conductivity $G$, and a low thermal conductivity $K$. But increasing the thermopower for bulk materials lead to a simultaneous decrease in the electrical conductivity due to Mott relation ($S\propto T[\partial \ln G(E)/\partial E]_{E=\epsilon_f} $) \cite{Mott1969,Jonson1980}. Wiedemann-Franz law ($K/GT=constant$) also prevents the high efficiency in bulk thermoelectric materials\cite{Jonson1980,Ashcroft1976}. So it is challenging to enhance thermoelectric efficiency for the bulk thermoelectric materials. Current bulk semiconductor thermoelectric materials have $ZT\approx1$ at room temperature, which corresponds to an efficiency about one sixth of the Carnot efficiency \cite{Benenti2017}. Hicks and Dresselhaus \cite{Hicks1993,Dresselhaus1993} pointed out that the low-dimensional materials exhibit better thermoelectric efficiency than their bulk counterpart. Mahan and Sofo \cite{Mahan1996} also predicted maximization of thermoelectric efficiency for materials with Dirac-delta-like density of states. Thus low-dimensional nano-materials are promising candidates for the thermoelectric power generation. These low-dimensional materials, such as molecular junctions, superlattice thin films, nanotubes, quantum wires, and quantum dots, provides a state of the art to manipulate the electron and phonon properties of the nano-system (for current review see \cite{Chen2012,Benenti2017}). The thermoelectric properties of these nanomaterials are strongly influenced by quantum confinement and Coulomb blockade effects which may lead to the failure of the Mott relation and also a violation of the Wiedemann-Franz law\cite{Boese2001,Lunde2006,Krawiec2006,Kubala2008,Murphy2008,Szczech2011}. Also, the phonon or lattice thermal conductance of low-dimensional systems is relatively small, which additionally contributes to the enhancement of thermoelectric efficiency \cite{Hicks1993,Hochbaum2008,Boukai2008,Markussen2009}.\\
There has been a significant progress in investigating the linear and non-linear thermoelectric transport properties of a quantum dot (QD) coupled to the normal metallic/ferromagnetic reservoirs\cite{Boese2001,Lunde2006,Krawiec2006,Kubala2008,Murphy2008,Szczech2011,Humphrey2002,Zianni2008,
Costi2010,Nakpathomkun2010,Sierra2014,Azema2014,Taylor2015,Bevilacqua2016,Erdman2017,Erdman2019,
Taniguchi2020,Svensson2013,Prete2019,Swirkowicz2009,Weymann2013}. For single-level QD coupled to normal reservoirs, the Coulomb interaction can enhance $ZT$ by suppressing the electron thermal conductance and increasing the thermopower\cite{Kubala2008,Taylor2015}. Also, $ZT$ is small in the low-temperature Kondo regime and quite high for the relatively larger temperatures or in the non-Kondo regime\cite{Krawiec2006}. It has also been found recently that, in the non-linear regime,  the applied thermal gradient and voltage biasing can manipulate the thermovoltage (electrical response to a temperature difference) and asymmetric heat dissipation, respectively, between the normal metallic reservoirs\cite{Sierra2014}.\\
When one of the reservoirs is a conventional s-wave Bardeen-Cooper-Schrieffer (BCS) superconductor (with superconducting energy gap $\Delta$), the Andreev tunnelling occurs in the normal metal-QD-superconductor interface in which an incident electron(hole) from the normal side is reflected as a hole(electron) and simultaneously creating(destroying) a Cooper pair in the superconductor\cite{Annett2004}. Hence, Andreev tunnelling leads to the formation of discrete Andreev bound states (ABS), with excitation energies within the superconducting energy gap. These ABS dominate the low-bias sub-gap electronic transport at low temperatures. The single particle or quasi-particle tunnelling becomes essential when the thermal energy is comparable to the superconducting energy gap or when the dot's energy level lies outside the superconducting energy gap.\\
Hybrid superconductor QD systems serve as a perfect platform to investigate the interplay between superconducting correlations and typical QD phenomena like Coulomb interaction and Kondo effect(for review see \cite{Martin2011}). The previous theoretical studies of the normal-QD-superconductor system (N-QD-S) deal with the equilibrium spectral properties\cite{Bauer2007,Sachin2020,Lim2020}, linear and non-linear transport under the pure electric response at very low temperature \cite{Fazio1998, Sun1999,Clerk2000,Sun2001,Cuevas2001,Krawiec2003,Tanaka2007,Domanski2008,Yamada2011}.\\
The thermoelectric properties of various single QD-superconductor-based hybrid systems such as N-QD-S \cite{Krawiec2008,Hwang2015}, F-QD-S (where, F stands for ferromagnetic reservoir) \cite{Hwang2016a,Hwang2016b,Hwang2017,Barnas2017} and S-QD-S \cite{Kleeorin2016,Kamp2019} have been rather weakly investigated in linear and non-linear transport regime. The thermoelectric properties of QD coupled to normal and superconducting reservoirs were first studied by Krawiec\cite{Krawiec2008} employing $U\rightarrow\infty$ Slave boson method.  In the linear response regime, he studied the background temperature dependence of thermoelectric quantities (electric and thermal conductance, thermopower, figure of merit, and Wiedemann-Franz ratio). Krawiec\cite{Krawiec2008} showed that superconductivity strongly modifies the thermal properties of the system, and suppression of the Andreev tunnelling due to strong on-site Coulom repulsion leads to a violation of the Wiedemann-Franz law (which indicates a non-Fermi liquid ground state).  Hwang et.al.\cite{Hwang2015} studied the N-QD-S system under the influence of the applied voltage as well as the temperature gradient in the non-linear regime by using the gauge-invariant non-linear thermoelectric transport theory. They showed that the $I-V$ characteristic of the N-QD-S system can be tuned by thermal gradient if the system is simultaneously voltage biased. Such cross effect occurs beyond linear response regime. These authors also proposed a highly efficient thermoelectric diode built from the coupling of a QD with a normal or ferromagnetic reservoir and a superconducting reservoir\cite{Hwang2016a}. Recently, the charge and spin thermoelectric effects in a QD coupled to ferromagnetic and superconducting reservoirs (F-QD-S) have been studied by few authors in the linear\cite{Hwang2016b,Barnas2017} and non-linear transport regime\cite{Hwang2016a,Hwang2017}.\\
 The heat transport and Peltier effect are well understood in the normal-metal-insulator-superconductor (NIS) micro-contact structure. For example, in reference {\cite{Bardas1995}} it was shown that for Andreev tunnelling dominate case, the heat flow is suppressed. This occurs due to particle-hole symmetry condition, i.e., electron and hole heat current compensate each other. On the other hand, for a correlated QD coupled between normal and superconducting reservoirs (N-QD-S), the magnitude and direction of Andreev and quasi-particle heat current can be manipulated by breaking the particle-hole symmetry, which is beyond the scope of the NIS situation. Particle-hole symmetry breaking can be achieved by tuning the QD energy level (applying external gate voltage) or applying an external magnetic field.\\
 As discussed above, the thermoelectric properties of N-QD-S and F-QD-S systems has been less studied\cite{Krawiec2008,Hwang2015,Hwang2016a,Hwang2016b,Hwang2017,Barnas2017}. Especially, the study of the heat transport (under pure voltage biasing) and thermoelectric particle-exchange heat engine beyond the linear regime is unexplored so far. Therefore, in this work, we provide a detail analysis of the low temperature electric and thermal response of a single-level quantum dot coupled to normal and BCS superconductor reservoirs (N-QD-S) within the Coulomb blockade regime by using Keldysh non-equilibrium Green's function technique\cite{Keldysh1965,Haug2008}. The assumption of QD with a single level is resonable if QD is small enough such that the energy levels separation between the ground and first excited state is much larger then the background thermal fluctuation ($\delta\epsilon>>k_BT$). Thus single level QD model is limited to the regime where quantum confinement effect dominates over thermal energy. The intradot Coulomb correlation is considered in the Hubbard-\Romannum{1} approximation\cite{Hubbard1963}. The electric and thermoelectric transport quantities i.e., the electrical conductance, thermal conductance, thermopower, and thermoelectric efficiency, are calculated in both linear and non-linear transport regime. We started from the linear response regime because the effect of tunnel coupling asymmetry, superconducting gap, and proximity induced local gap on the properties of thermoelectric heat engine are missing from earlier studies and needed further analysis. The non-linear regime is investigated for two different cases : (\romannum{1}) voltage-driven case i.e. voltage biasing without temperature gradient, (\romannum{2}) temperature driven case i.e. when N-QD-S system works as a thermoelectric particle-exchange heat engine or power generator.\\
\begin{figure}[!htb]
\includegraphics
  [width=0.85\hsize]
{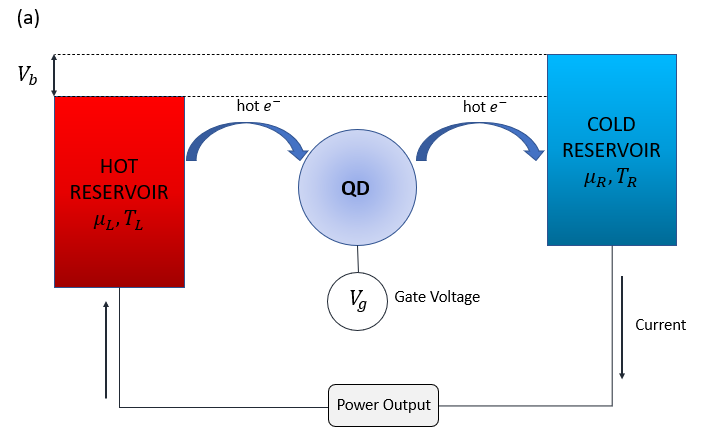}
\end{figure}
\begin{figure}[!htb]
\includegraphics
  [width=0.85\hsize]
{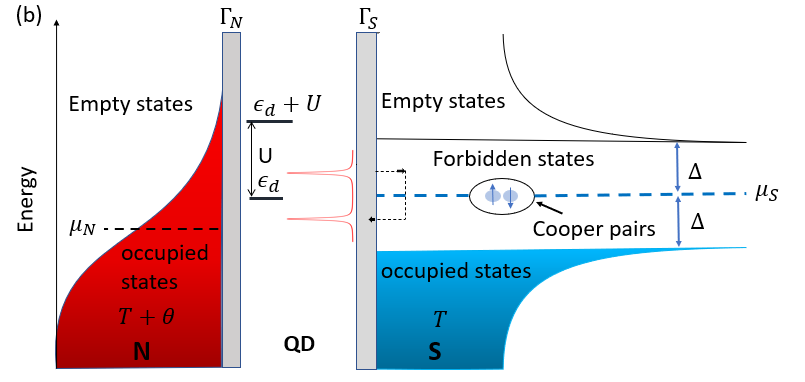}
\caption {(a): A schematic diagram of the thermoelectric particle-exchange heat engine based on the QD. The engine operates by a continuous flow of electrons between hot and cold reservoirs. The electrons carry charge and energy as they  flow through the QD system and they do work against the electric field created by charge imbalance. (b): The corresponding schematic electronic band diagram when the hybrid N-QD-S system work as a thermoelectric particle-exchange heat engine. The single level QD with two effective levels at $\epsilon_d$ and $\epsilon_d+U$ (due to finite Coulomb interaction effect) is connected to hot normal and cold superconducting reservoirs with applied bias $\mu_S-\mu_N=eV$. Where $\mu_N$ and $\mu_S$ are the chemical potentials of the normal and superconducting reservoirs respectively.}
\label{fig:1}
\end{figure}
Fig.\ref{fig:1}(a) illustrates the schematic diagram of the thermoelectric heat engine based on QD. The corresponding band diagram of the N-QD-S system working as a thermoelectric heat engine is shown in Fig.\ref{fig:1}(b). When a thermal gradient is applied ($T_N-T_S=\theta$), the electrons flows from the left normal reservoir to the right superconducting reservoir and creates a potential difference $\mu_N-\mu_S=eV_{th}$, where $V_{th}$ is thermovoltage. The N-QD-S works as a thermoelectric heat engine when the temperature field on the electron motion overcomes the electric field on them, i.e a reverse voltage $\mu_S-\mu_N=eV$ must be applied.\\
This paper is organized as follows : A detailed discussion of the model Hamiltonian and theoretical formalism is provided in the preceding section  \hyperref[sec:Model]{\Romannum{2}}. The numerical results and discussion for linear and non-linear regime are given in section \hyperref[sec:Result]{\Romannum{3}}. Section \hyperref[sec:Conclusion]{\Romannum{4}} concludes the present work.
\section{Model Hamiltonian and theoretical formalism}\label{sec:Model}
To analyse the thermoelectric transport properties, we describe the N-QD-S system by the following Anderson+BCS model Hamiltonian in second quantization,
\begin{equation}\label{eq:1}
  \begin{aligned}
\hat{H}=\hat{H}_{N}+\hat{H}_{S}+\hat{H}_{QD}+\hat{H}_{T}
 \end{aligned}
\end{equation}
where
\begin{equation*}
\begin{aligned}
& \hat{H}_{N} = \sum_{k\sigma}(\epsilon_{k,N}c^\dagger_{k\sigma,N}c_{k\sigma,N}),\\
& \hat{H}_{S} = \sum_{k\sigma}(\epsilon_{k,S}c^\dagger_{k\sigma,S}c_{k\sigma,S})+
\sum_{k}\left(\Delta c^\dagger_{k\uparrow,S}c^\dagger_{-k\downarrow,S}+H.c\right),\\
& \hat{H}_{QD} = \sum_{\sigma} \epsilon_{d}n_{\sigma}+Un_{\uparrow}n_{\downarrow},\\
& \hat{H}_{T} = \sum_{k\sigma,\alpha\in N,S}(V_{k,\alpha}d^\dagger_{\sigma}c_{k\sigma\alpha}+{V^\ast_{k,\alpha}}c^\dagger_{k\sigma,\alpha}d_{\sigma}).\\
\end{aligned}
\end{equation*}
$\hat{H}_{N}$  describes the normal metallic  reservoir in the non-interacting quasi-particle approximation with   single electron kinetic energy $\epsilon_{k,N}$ and $c_{k\sigma,N}(c^\dagger_{k\sigma,N})$ is the annihilation(creation) operator of an electron with spin $\sigma$ and wave vector $\vec{k}$.\\
$\hat{H}_{S}$ describes the superconducting reservoir. $c_{k\sigma,S}(c^\dagger_{k\sigma,S})$ is the annihilation(creation) operator of an electron with spin $\sigma$, wave vector $\vec{k}$ and energy $\epsilon_{k,S}$. The second term in $\hat{H}_{S}$ describes the BCS pair interaction, with a superconducting energy gap $\Delta$.\\
$\hat{H}_{QD}$  describes the Hamiltonian for single-level QD with energy $\epsilon_d$, and $d_\sigma(d^\dagger_\sigma)$ is the annihilation(creation) operator of electron with spin $\sigma$ on the QD and $n_{\sigma}=d_\sigma^\dagger d_\sigma$ is number operator. The QD can have maximum occupancy of two electrons with opposite spins. We also consider the intradot electron-electron Coulomb repulsion with the interaction strength U.\\
$\hat{H}_{T}$ represents the tunnelling Hamiltonian between the QD energy level and reservoirs with $V_{k\alpha}$ as the tunnelling amplitude between the QD and the $\alpha$-reservoir ($\alpha\in N,S$).\\
To diagonalized the BCS part of the Hamiltonian, we use Bogoliubov transformation method which defines the new Fermionic quasi-particle operator $\gamma_{k\sigma}(\gamma^\dagger_{k\sigma})$ and coefficients $u_k$ and $v_k$ 
\begin{equation}
\begin{aligned}
c_{k\uparrow,S} = u^\ast_k\gamma_{k\uparrow}+v_k\gamma^\dagger_{-k\downarrow},
\;\;
c^\dagger_{-k\downarrow,S} = u_k\gamma^\dagger_{-k\downarrow}-v^\ast_k\gamma_{k\uparrow}
\end{aligned}
\end{equation}
with normalization condition $|{u_k}|^2+|{v_k}|^2=1$. 
Substituting above equation in Eq.\eqref{eq:1} yields following effective model Hamiltonian
\begin{equation} \label{eq:3}
\begin{aligned}
\hat{H} = & \sum_{k,\sigma}(\epsilon_{k,N}c^\dagger_{k\sigma,N}c_{k\sigma,N})
+\sum_{k\sigma}(V_{k,N}d^\dagger_{\sigma}c_{k\sigma,N}+H.c)+\\
& \sum_{k,\sigma}(E_{k}\gamma^\dagger_{k\sigma}\gamma_{k\sigma})
+\sum_{k\sigma}(V_{k,S}u^\ast_kd^\dagger_{\sigma}\gamma_{k\sigma}+H.c)+\\
& \sum_{k}[V^\ast_{k,S}v_k(d^\dagger_{\uparrow}\gamma^\dagger_{-k\downarrow}-d^\dagger_{\downarrow}\gamma^\dagger_{k\uparrow})+ H.c]+\\
& \sum_{\sigma} \epsilon_{d}n_{\sigma}+Un_{\uparrow}n_{\downarrow}
\end{aligned}
\end{equation}
where $E_k = \sqrt{\epsilon^2_{k,S}+|\Delta|^2}$ is the excitation quasi-particle energy of the superconducting reservoir. The coefficients $u_k$ and $v_k$ read
\begin{equation}
|{u_k}|^2=\frac{1}{2}\left(1+\frac{\epsilon_{k,S}}{\sqrt{{\epsilon^2_{k,S}}+|\Delta|^2}}\right)
\end{equation}
\begin{equation}
 |{v_k}|^2=\frac{1}{2}\left(1-\frac{\epsilon_{k,S}}{\sqrt{{\epsilon^2_{k,S}}+|\Delta|^2}}\right)
\end{equation}
To solve above effective Hamiltonian (Eq.\eqref{eq:3}), we use the Green's function equation of motion method with Zubarev notation \cite{Zubarev1960} for the retarded Green's function $G^r_{A,B}(t)={\langle\langle{\hat{A}|\hat{B}}\rangle\rangle}=-i\theta(t)\langle[A(t),B(0)]_+\rangle$, where $\hat{A}$ and $\hat{B}$ are creation or annihilation operators, $\theta(t)$ is unit step or heaviside function and $[\hat{A},\hat{B}]_{\pm}=\hat{A}\hat{B}\pm \hat{B}\hat{A}$. The Fourier transform of the above retarded Green's function ${\langle\langle{\hat{A}|\hat{B}}\rangle\rangle}_{\omega}$ must satisfies the equation of motion (EOM),
\begin{equation}
\begin{aligned} 
\omega \langle\langle{\hat{A}|\hat{B}}\rangle\rangle_\omega=\langle{[\hat{A},\hat{B}]_+}\rangle+\langle\langle{[\hat{A},\hat{H}]_{-}|\hat{B}}\rangle\rangle_\omega
\end{aligned}
\end{equation}
For non-zero Coulomb correlation ($U\neq0$) an equation of motion for a given Green's function involves higher-order coupled Green's functions, thus creating a hierarchy of equations of motion (EOM). In order to truncate the hierarchy of equations one need a decoupling scheme for higher order Green's functions and maintain self-consistency.\\
In Nambu representation, we define the retarded Green's function of the QD as a $2\times 2$ matrices
\begin{widetext}
 \begin{equation} \label{eq:7}
 {\bf{G}}^{r}_{d}(\omega)={\left\langle\left\langle{
\begin{pmatrix}
d_{\uparrow}\\
d_{\downarrow}^{\dagger}\\
\end{pmatrix}
\begin{pmatrix}
d_{\uparrow}^{\dagger} & d_{\downarrow}
\end{pmatrix}
}\right\rangle\right\rangle}_{\omega}=
\begin{pmatrix}
 \langle\langle{d_{\uparrow}|d_{\uparrow}^{\dagger}}\rangle\rangle_{\omega} & \langle\langle{d_{\uparrow}|d_{\downarrow}}\rangle\rangle_{\omega} \\
 \langle\langle{d_{\downarrow}^{\dagger}|d_{\uparrow}^{\dagger}}\rangle\rangle_{\omega} & \langle\langle{d_{\downarrow}^{\dagger}|d_{\downarrow}}\rangle\rangle_{\omega} \\
\end{pmatrix}=
\begin{pmatrix}
 G^r_{d,11}(\omega) & G^r_{d,12}(\omega) \\
 G^r_{d,21}(\omega) & G^r_{d,22}(\omega)\rangle\rangle \\
\end{pmatrix}
\end{equation}
\end{widetext}
Where the diagonal components of ${\bf{G}}^{r}_{d}(\omega)$ represents the single particle retarded Green's function of electron with spin $\sigma=\uparrow$ and hole with spin $\sigma=\downarrow$  respectively. The off-diagonal component represents the superconducting paring correlation on the QD.\\
By evaluating different commutator and anti-commutator brackets we drive the following EOM for the single electron Green's function with spin $\sigma=\uparrow$\\
\begin{equation}
\begin{aligned} 
& (\omega-\epsilon_d)\langle\langle{d_{\uparrow}|d_{\uparrow}^{\dagger}}\rangle\rangle_{\omega}=1
+\sum_kV^{\ast}_{k,N}\langle\langle{c_{k\uparrow,N}|d_{\uparrow}^{\dagger}}\rangle\rangle_{\omega}+\\
& \sum_kV_{k,S}u^{\ast}_k\langle\langle{\gamma_{k\uparrow}|d_{\uparrow}^{\dagger}}\rangle\rangle_{\omega}
+\sum_kV_{k,S}v_k\langle\langle{\gamma_{-k\downarrow}|d_{\uparrow}^{\dagger}}\rangle\rangle_{\omega}+\\
& U\langle\langle{d_{\uparrow}d_{\downarrow}^{\dagger}d_{\downarrow}|d_{\uparrow}^{\dagger}}\rangle\rangle_{\omega}.
\end{aligned} 
\end{equation}
Similarly, one may write down the equation of motion for other Green's functions in Eq.\eqref{eq:7} and for the relevant correlation functions appearing in those equation of motions.\\
We treated the Coulomb correlations within Hubbard-\Romannum{1} approximation, which correctly describes the Coulomb blockade effects\cite{Hubbard1963, Rozhkov2010,Vovchenko2014}. It is important to pointed out that in the Hubbard-\Romannum{1} decoupling scheme one can manage to retain higher-order tunneling processes through QD as a manifestation  of electron-electron interaction, which are missing in the Hartree-Fock approximation based on weak electronic correlation. Within Hubbard-\Romannum{1} decoupling scheme EOM for higher order Green's functions of the form $U\langle\langle{d_{\uparrow} d_{\downarrow}^{\dagger}d_{\downarrow}|d_{\uparrow}^{\dagger}}\rangle\rangle_{\omega}$ is simplified by using,
\begin{equation}
U\langle\langle{c_{k\pm\sigma,N} d_{-\sigma}^{\dagger}d_{-\sigma}|d_{+\sigma}^{\dagger}}\rangle\rangle_{\omega}\rightarrow U\langle n_{-\sigma}\rangle\langle\langle{c_{k\pm\sigma,N}|d_{+\sigma}^{\dagger}}\rangle\rangle_{\omega}
\end{equation}
\begin{equation}
U\langle\langle{\gamma_{k\pm\sigma} d_{-\sigma}^{\dagger}d_{-\sigma}|d_{+\sigma}^{\dagger}}\rangle\rangle_{\omega}\rightarrow U\langle n_{-\sigma}\rangle\langle\langle{\gamma_{k\pm\sigma}|d_{+\sigma}^{\dagger}}\rangle\rangle_{\omega}
\end{equation}
where $\pm\sigma \in \uparrow,\downarrow$ and $\langle{n_{-\sigma}}\rangle=\langle d_{-\sigma}^{\dagger}d_{-\sigma}\rangle$ denotes the quantum statistical average value of occupation number with spin $-\sigma$.\\
Within Hubbard-\Romannum{1} decoupling scheme the correlations involving reservoir electrons in the higher order Green's function and the spin flip processes on the dot are neglected (i.e  $\langle{d_{-\sigma}^{\dagger}d_{+\sigma}}\rangle$=0). Thus the formalism and results explained in the present paper are relevant for temperatures higher than the temperature associated with the Kondo effect (i.e., Kondo temperature $T_K$)\cite{Yamada2011}. We also assume that the Coulomb correlation effects are manifested only in the diagonal elements of the Green's function (Eq.\eqref{eq:7}) i.e. self-consistent determination of the proximity induced local superconducting gap or pairing amplitude on the QD site is excluded from our analysis (i.e., $\langle d_{\uparrow}d_{\downarrow}\rangle\rightarrow0$), which drastically reduces the computational time. This assumption is justified in subsection \hyperref[sec:induced superconducting gap] {\Romannum{3}.D}, where we have shown that the inclusion of self-consistent equation for $\langle d_{\uparrow}d_{\downarrow}\rangle$ does not affect the electric and thermoelectric transport properties analysed in the present work.\\
For simplification the tunnelling amplitude is considered $k$ independent i.e $V_{k,\alpha}=V_{\alpha}$ for $V_{k,\alpha}<<D$ (wide band), where $-D\leq\epsilon_{k,\alpha}\leq D$, with $D$ as the half bandwidth. The tunneling coupling strength of the QD to the $\alpha$-reservoir ($\alpha\in N,S$) is defined by $\Gamma_{\alpha}=2\pi|V_{\alpha}|^2\rho_{0,\alpha}$, where normal metallic density of states $\rho_{0,\alpha}$ is constant in the range of energy around Fermi level (flat band). Here $\rho_{0,S}$ is modified in the superconducting state.\\
Finally after solving coupled EOM based on above Hubbard-\Romannum{1} scheme we arrive at the expression for the retarded Green's function of electron with spin $\sigma=\uparrow$ and off-diagonal superconducting pairing correlation on the QD,
\begin{widetext}
\begin{equation}
\resizebox{0.95\hsize}{!}{$G_{d,11}^{r}(\omega)=\langle\langle{d_{\uparrow}|d_{\uparrow}^{\dagger}}\rangle\rangle=\cfrac{\left(1+\cfrac{U\langle{n_{\downarrow}}\rangle}{\omega-\epsilon_d-U}\right)}{\left[\omega-\epsilon_d+\left(\cfrac{i\Gamma_N}{2}+\beta(\omega)\right)\left(1+\cfrac{U\langle{n_{\downarrow}}\rangle}{\omega-\epsilon_d-U}\right)-\cfrac{\left(1+\cfrac{U\langle{n_{\downarrow}}\rangle}{\omega-\epsilon_d-U}\right)\left(1-\cfrac{U\langle{n_{\uparrow}}\rangle}{\omega+\epsilon_d+U}\right)\left(\cfrac{\Delta}{|\omega|}\beta(\omega)\right)^2}{\omega+\epsilon_d+\left(\cfrac{i\Gamma_N}{2}+\beta(\omega)\right)\left(1-\cfrac{U\langle{n_{\uparrow}}\rangle}{\omega+\epsilon_d+U}\right)}\right]}$}
\end{equation}
\begin{equation}
G_{d,21}^{r}(\omega)=\langle\langle{d_{\downarrow}^{\dagger}|d_{\uparrow}^{\dagger}}\rangle\rangle=\left[\cfrac{\left(1-\cfrac{U\langle{n_{\downarrow}}\rangle}{\omega+\epsilon_d+U}\right)\left(\cfrac{\Delta}{|\omega|}\beta(\omega)\right)}{\omega+\epsilon_d+\left(\cfrac{i\Gamma_N}{2}+\beta(\omega)\right)\left(1-\cfrac{U\langle{n_{\downarrow}}\rangle}{\omega+\epsilon_d+U}\right)}\right]\times G_{d,11}^{r}(\omega)
\end{equation}
with
\begin{equation}
\beta(\omega)=\cfrac{\Gamma_S}{2}\rho_S(\omega)=\cfrac{\cfrac{\Gamma_S}{2}\omega}{\sqrt{\Delta^2-\omega^2}}\theta(\Delta-|\omega|)+\cfrac{\cfrac{i\Gamma_S}{2}|\omega|}{\sqrt{\omega^2-\Delta^2}}\theta(|\omega|-\Delta)
\end{equation}
\end{widetext}
where $\rho_S$ is the modified BCS density of states.\\
The other matrix elements is given by $G_{d,22}^{r}(\omega)=-G_{d,11}^{r}(-\omega)^{\ast}$ and $G_{d,12}^{r}(\omega)=G_{d,21}^{r}(-\omega)^{\ast}$.
These retarded Green's functions allow us to calculate the advanced and lesser/greater Green's functions and eventually the single particle thermoelectric properties. The  averaged occupation of electrons per spin on the quantum dot ($\langle{n_{\uparrow}}\rangle$=$\langle{n_{\downarrow}}\rangle$ for non-magnetic system) is calculated using the self-consistent integral equation of the form
\begin{equation} \label{eq:14}
\langle{n_{\sigma}}\rangle=\frac{-i}{2\pi}\int^{\infty}_{-\infty}G^{<}_{d,11}(\omega) d\omega
\end{equation}
where the lesser Green's function $G^{<}_{d}$ is introduced which obeys the Keldysh equation (in the matrix form) \cite{Haug2008},
 \begin{equation} \label{eq:15}
{\bf{G}}^{<}_{d}(\omega)=-{\bf{G}}^{r}_{d}(\omega){\bf{\Sigma}}^{<}_{d}(\omega){\bf{G}}^{a}_{d}(\omega)
\end{equation}
where ${\bf{G}}^{a}_{d}(\omega)=\left[{\bf{G}}^{r}_{d}(\omega)\right]^{\dagger}$ is the advanced Green's function and ${\bf{\Sigma}}^{<}_{d}(\omega)= -\sum_{\alpha\in N,S}\left[{\bf{\Sigma}}^{r}_{\alpha}-{\bf{\Sigma}}^{a}_{\alpha}\right]{\bf{f}}_{\alpha}(\omega)$ is the lesser self energy matrix given by,
\begin{widetext}
\begin{equation}
\resizebox{0.95\hsize}{!}{${\bf{\Sigma}}^{<}_{d}(\omega)=
\begin{pmatrix}
-i\Gamma_{N}f_N(\omega-\mu_N)-\cfrac{i\Gamma_S|\omega|}{\sqrt{\omega^2-\Delta^2}} \theta(|\omega|-\Delta)f_S(\omega-\mu_S) & \cfrac{i\Gamma_{S}\Delta}{\sqrt{\omega^2-\Delta^2}}\theta(|\omega|-\Delta)f_S(\omega-\mu_S)\\
 \cfrac{i\Gamma_{S}\Delta}{\sqrt{\omega^2-\Delta^2}}\theta(|\omega|-\Delta)f_S(\omega-\mu_S) & -i\Gamma_{N}f_N(\omega+\mu_N)-\cfrac{i\Gamma_S|\omega|}{\sqrt{\omega^2-\Delta^2}} \theta(|\omega|-\Delta)f_S(\omega-\mu_S)
\end{pmatrix}$}
\end{equation}
Thus the lesser Green's function for electrons on the QD is given by,
\begin{equation}
\begin{aligned}
G^{<}_{d,11}(\omega) = & i\Gamma_{N}f_N(\omega-\mu_N)|G_{d,11}^{r}(\omega)|^2+
i\Gamma_{N}f_N(\omega+\mu_N)|G_{d,12}^{r}(\omega)|^2+\\
& \cfrac{i\Gamma_S|\omega|}{\sqrt{\omega^2-\Delta_2}}\;\theta(|\omega|-\Delta)f_S(\omega-\mu_S)\left[|G_{d,11}^{r}(\omega)|^2+|G_{d,12}^{r}(\omega)|^2
-\cfrac{2\Delta}{|\omega|} Re\left(G_{d,11}^{r}(\omega).G_{d,12}^{a}(\omega)\right)\right]
\end{aligned}
\end{equation}
\end{widetext}
where $\theta(|\omega|-\Delta)$ is the unit step function and $f_{\alpha\in N,S}(\omega\mp\mu_{\alpha})=\left[{exp((\omega\mp\mu_{\alpha})/k_BT_{\alpha})+1}\right]^{-1}$ is the Fermi-Dirac distribution function of reservoirs with temperature $T_{\alpha}$ and chemical potential $\pm\mu_{\alpha}$ (measured from Fermi level $\epsilon_f$ or $\mu_f=0$).\\
In the linear response regime, i.e., for  small Voltage biasing ($\mu_S-\mu_N=e\delta V\rightarrow 0$) and small temperature gradients ($T_N-T_S=\delta\theta\rightarrow 0$) between the reservoirs, the Fermi function of the normal and the superconducting reservoirs can be expanded around the equilibrium value (average $T$ with $\mu_f=0$), which gives
\begin{equation}
f_{\alpha}(\omega\pm\mu_{\alpha}) \approx f_{eq}\pm\cfrac{df_{eq}}{d\omega}\left[\mu_{\alpha}-\left(\cfrac{\omega}{T}\right)(T-T_{\alpha})\right]
\end{equation}
where $f_{eq}\!=\!\left[exp(\omega/k_BT)+1\right]^{-1}$ is the equilibrium Fermi-Dirac distribution function with $\mu_S=\mu_N$ and $T_S=T_N$.\\
Substituting above linear relation into Eqs.\eqref{eq:22}-\eqref{eq:27} (given below) gives
the electrical current and heat current satisfying the Onsager relation\cite{Mahan2000},
  \begin{equation}\label{eq:16}
  \begin{pmatrix}
 I_C \\
 \\
 J_Q \\
 \\
 \end{pmatrix}
 =
  \begin{pmatrix}
 e^2L_0 & \cfrac{e}{T}L_1 \\
 \\
 eL_1 & \cfrac{1}{T}L_2 \\
 \end{pmatrix}
 \\
  \begin{pmatrix}
 \delta V \\
 \\
  \delta\theta \\
  \\
 \end{pmatrix}
\end{equation}
with thermoelectric response functions
\begin{equation*}
\begin{aligned}
L_0 &= \cfrac{2}{h}\int{\left(\cfrac{-df_{eq}}{d\omega}\right) (2T_A(\omega)+T_{QP}(\omega))d\omega}\\
L_1 &= \cfrac{2}{h}\int{\omega\left(\cfrac{-df_{eq}}{d\omega}\right) T_{QP}(\omega)d\omega}\\
L_2 &= \cfrac{2}{h}\int{\omega^2\left(\cfrac{-df_{eq}}{d\omega}\right) T_{QP}(\omega)d\omega}\\
\end{aligned}
\end{equation*}
Here, $e$ and $h$ denote the magnitude of the electronic charge and Planck’s constant, respectively.\\
\\
$T_A(\omega)=\Gamma_N^2 |G_{d,12}^r(\omega)|^2$ is the Andreev tunnelling amplitude and $T_{QP}(\omega) =\cfrac{\Gamma_N\Gamma_S|\omega|}{\sqrt{\omega^2-\Delta^2}}\;\theta(|\omega|-\Delta)\times
\left[|G_{d,11}^r(\omega)|^2+|G_{d,21}^r(\omega)|^2-\cfrac{2\Delta}{|\omega|}Re{(G_{d,11}^r(\omega)\;G_{d,12}^a(\omega))}\right]$
is the quasi-particle tunnelling amplitude.\\ 
The thermoelectric transport quantities (electrical conductance $G$, thermopower or Seeback coefficient $S$, and electronic contribution to thermal conductance $K$) are then obtained from Eq.\eqref{eq:16}.
\begin{align}
\label{eq:17}
& G = \lim_{\delta V \to 0} {\cfrac{I_C}{\delta V}}\biggr\rvert_{\delta\theta=0} = e^2L_0\\
\label{eq:18}
& S = \lim_{\delta\theta\to 0} {\cfrac{\delta V}{\delta\theta}}\biggr\rvert_{I_C=0} = -\cfrac{1}{eT}\frac{L_1}{L_0}\\
\label{eq:19}
& K = \lim_{\delta\theta \to 0} {\cfrac{J_Q}{\delta\theta}}\biggr\rvert_{I_C=0} = \cfrac{1}{T}\left[L_2-\cfrac{L_1^2}{L_0}\right]=\cfrac{1}{T}L_2-S^2GT
\end{align}
The electrical conductance is defined as the flow of charge current per unit voltage between the isothermal reservoirs ($\delta\theta=0$). The thermopower (Seebeck coefficient) is defined as the generated voltage per unit thermal gradients in open circuit condition i.e. $I_C=0$. The electronic thermal conductance is usually given by the heat flow through the central region (in present case QD) when it is coupled between electrically insulating source and drain reservoirs at different temperatures. In such systems, the reservoirs impose open circuit condition $I_C = 0$. However, for a thermoelectric system with non-zero $L_1$ or $S$, a voltage will build up across the QD proportional to the temperature difference. Thus, total electronic thermal conductance $K$ is reduced as compared to electrically insulating reservoirs by a factor of $S^2GT$. Where $P=S^2G$ is the corresponding power factor.\\
The performance of the thermoelectric heat engine in the linear response regime is determined by a dimensionless thermoelectric figure of merit $ZT$.
\begin{equation}\label{eq:20}
ZT = \cfrac{S^2GT}{K}=\cfrac{S^2GT}{K_e+K_{ph}}
\end{equation}
Since we are interested in understanding the electronic thermal properties at low temperatures, we only consider the thermal contribution by electrons, and the lattice or phonon thermal contribution is negligible at low temperatures($K_e\approx K$)\cite{Hwang2016b,Yang2020}.\\
In the linear response regime, the relationship between figure of merit $ZT$ and efficiency at maximum power output $\eta_{P_{max}}$ is given by\cite{Erdman2017,Benenti2017}
\begin{equation} \label{eq:21}
\eta_{P_{max}} = \cfrac{\eta_C}{2}\;\cfrac{ZT}{ZT+2}
\end{equation}
where $\eta_C$ is the Carnot efficiency.\\
In the non-linear regime the system is under the influence of finite voltage biasing $\mu_N-\mu_S=eV$ (say $\mu_N=eV$ and $\mu_S=0$) and/or temperature gradient $T_N-T_S=\theta$ (say $T_N=T+\theta$ and $T_S=T$). Thus the definition of linear response regime fails, and one needs to go beyond this limit. In the non-equilibrium steady-state condition, the net current flowing through the left normal and right superconducting reservoirs is $I_N=-I_S\equiv I_C$ (current conservation) and can be evaluated from the time evolution of the occupation number operator of the left normal reservoir ($d\langle{-e\sum_{k}c^{\dagger}_{k,N}c_{k,N}}\rangle/dt$)\cite{Haug2008, Meir1992}.
\begin{equation}\label{eq:22}
I_N\equiv I_C = I_A+I_{QP};
\end{equation}
where
\begin{equation}\label{eq:23}
\resizebox{0.9\hsize}{!}{$I_A = \cfrac{2e}{h} \int{\left[f_N(\omega-\mu_N)-f_N(\omega+\mu_N)\right] T_A(\omega)\;d\omega}$}
\end{equation}
\begin{equation}\label{eq:24}
\resizebox{0.9\hsize}{!}{$I_{QP} = \cfrac{2e}{h} \int{\left[f_N(\omega-\mu_N)-f_S(\omega-\mu_S)\right] T_{QP}(\omega)\;d\omega}$}
\end{equation}
are Andreev and quasi-particle contribution to electrical/charge current respectively.\\
The heat current is evaluated from the rate of energy flow at the normal reservoir side and Joule heating in the presence of voltage biasing.
\begin{equation}\label{eq:25}
J_N\equiv J_Q = \cfrac{-i}{\hbar}\langle{[\hat{H},\hat{H}_N]}\rangle-\cfrac{\mu_NI_C}{e}= J_A+J_{QP};
\end{equation}
where
\begin{equation}\label{eq:26}
\resizebox{0.9\hsize}{!}{$J_A =-4\mu_N\cfrac{1}{h} \int{\left[f_N(\omega-\mu_N)-f_N(\omega+\mu_N)\right] T_A(\omega)\;d\omega=-2\cfrac{\mu_NI_A}{e}}$}
\end{equation}
\begin{equation}\label{eq:27}
\resizebox{0.9\hsize}{!}{$J_{QP} = \cfrac{2}{h} \int{(\omega-\mu_N)\left[f_N(\omega-\mu_N)-f_S(\omega-\mu_S)\right] T_{QP}(\omega)\;d\omega}$}
\end{equation}
are Andreev and quasi-particle contribution to heat current respectively. The heat current satisfy the condition $J_N+J_S=-I_C(\mu_N-\mu_S)/e$ and we have shown the calculation for heat current flowing at the normal reservoir side i.e. $J_Q\equiv J_N=-J_S-I_C(\mu_N-\mu_S)/e$.\\
Also, the prefactor 2 in charge and heat current is due to the spin degeneracy. Note that for thermal gradient without voltage biasing (i.e. $\mu_N=eV=0$) the subgap ABS heat current is zero and only quasi-particle contributes to the heat current.\\
In order to use N-QD-S as a heat engine or power generator, the temperature gradient $\theta$ is set larger then zero. Due to this temperature difference electrons move from left reservoir to the right reservoir and thus create a potential difference ($\mu_N-\mu_S = eV_{th}$) due to accumulation of electrons on the right reservoir and positive charge to the left reservoir.\\
The thermovoltage ($V_{th}$) thermopower ($S$) and electronic thermal conductance ($K$) is determined from the open circuit condition,
\begin{equation} \label{eq:28}
I_C(V_{th},\theta) = I_A(V_{th},\theta)+I_{QP}(V_{th},\theta)=0
\end{equation}
For finite Voltage biasing and temperature gradient above equation is solved numerically to obtain $V_{th}$ and eventually $S=\cfrac{V_{th}}{\theta}$ and $K = \cfrac{J_Q}{\theta}$.\\
\\
The heat engine generates a finite power  $P$ between $V = 0$ and $V = V_{th}$ and is given by,
\begin{equation}\label{eq:29}
 P=-I_CV
\end{equation}
where $V=(\mu_S-\mu_N)/e$ is bias voltage applied to counteract the thermally induced current i.e power is generated when current is driven against the potential difference.\\
The thermoelectric efficiency is defined as the  ratio between the generated output power (nonlinear current times voltage) and the nonlinear input heat current i.e $\eta = output\;power/input\;heat = P/J_Q$. The maximal power generated by the heat engine is calculated numerically by using Eq.\eqref{eq:28}\&\eqref{eq:29} and the relative efficiency at maximal power output is given by,
\begin{equation}
 \left(\cfrac{\eta_{P_{max}}}{\eta_C}\right) = \cfrac{P_{max}}{J_Q}\times{\cfrac{T+\theta}{\theta}}
\end{equation}
where Carnot efficiency $\eta_C=\cfrac{\theta}{T+\theta}$.
\section{Result and discussion}
\label{sec:Result}
This section presents the numerical results obtained using MATLAB for the linear (subsection \hyperref[sec:Linear]{\Romannum{3}.A}) and non-linear response under electric and temperature field in the Coulomb blockade regime. We analyse two different situations in the non-linear transport regime. In the first situation (subsection \hyperref[sec:Voltage Driven]{\Romannum{3}.B}), we consider a voltage-driven case for isothermal reservoirs ($\theta=0$) and discuss the Andreev and quasi-particle charge and heat transport.  In the second situation (subsection \hyperref[sec:Heat Engine]{\Romannum{3}.C}), the  N-QD-S system is discussed in the context of thermoelectric heat engine for finite $\theta$ and $V$.  In our calculations, all energies are expressed in the unit of superconducting energy gap $\Delta$.
\subsection{Linear Response regime}\label{sec:Linear}
\begin{figure*}[!htb]
\includegraphics
  [width=0.85\hsize]
  {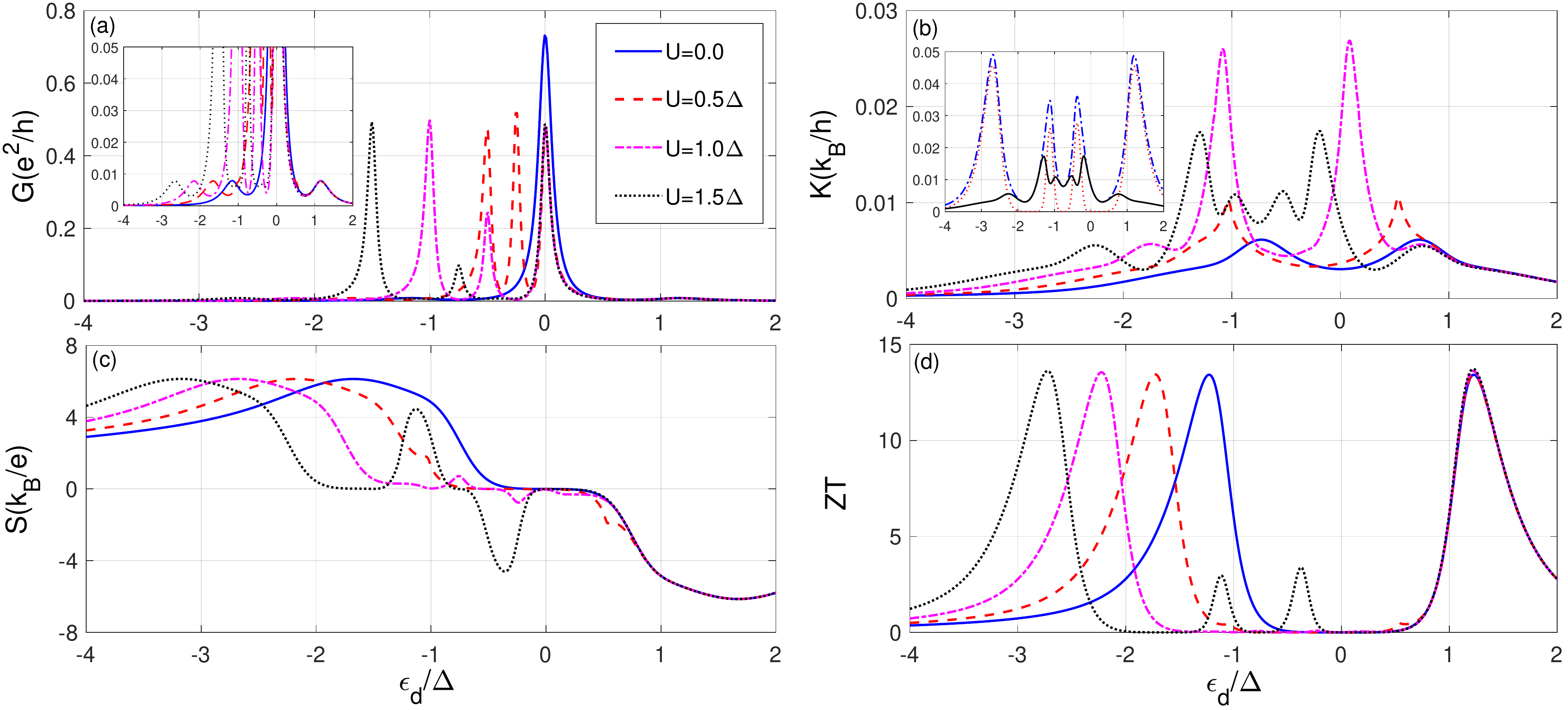}
\caption {Variation of  electrical conductance $G$, thermal conductance $K$, thermopower $S$, and figure of merit $ZT$ with the quantum dot energy level $\epsilon_d$ in the linear response regime for different values of the on-site Coulomb repulsion $U$. The other parameters are : $\Gamma_S=\Gamma_N=0.1\Delta$ and $k_BT=0.2\Delta$. The inset in (a) shows the close-up view of the linear electrical conductance $G$. The inset in (b) shows $L_2/T$ (blue dash-dot line), $S^2GT$ (red dotted line) and net electronic thermal conductance $K$ (black solid line) as a function of QD energy level $\epsilon_d$ for $U=1.5\Delta$ and other parameters remains same.}
\label{fig:2}
\end{figure*}
\begin{figure*}[!htb]
\includegraphics
  [width=0.85\hsize]
  {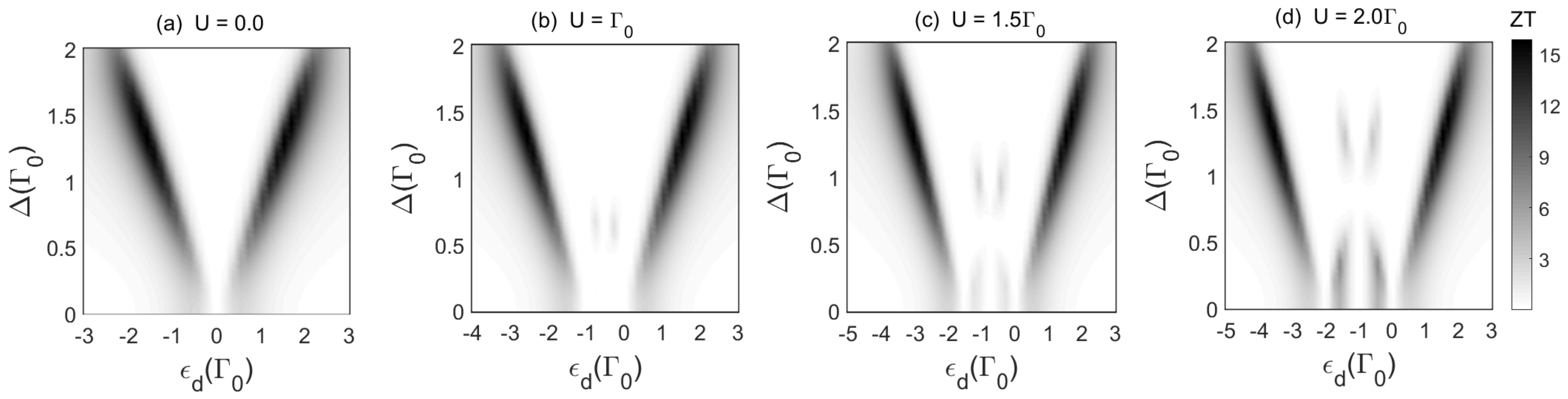}
\caption {Variation of the figure of merit $ZT$ with the QD energy level $\epsilon_d$ and superconducting gap $\Delta$ in the linear response regime for different on-site Coulomb repulsion $U$ with $\Gamma_S=\Gamma_N=0.1\Gamma_0$, $k_BT=0.2\Gamma_0$ and $\Gamma_0$ is energy unit.}
\label{fig:3}
\end{figure*}
\begin{figure*}[!htb]
\includegraphics
  [width=0.85\hsize]
  {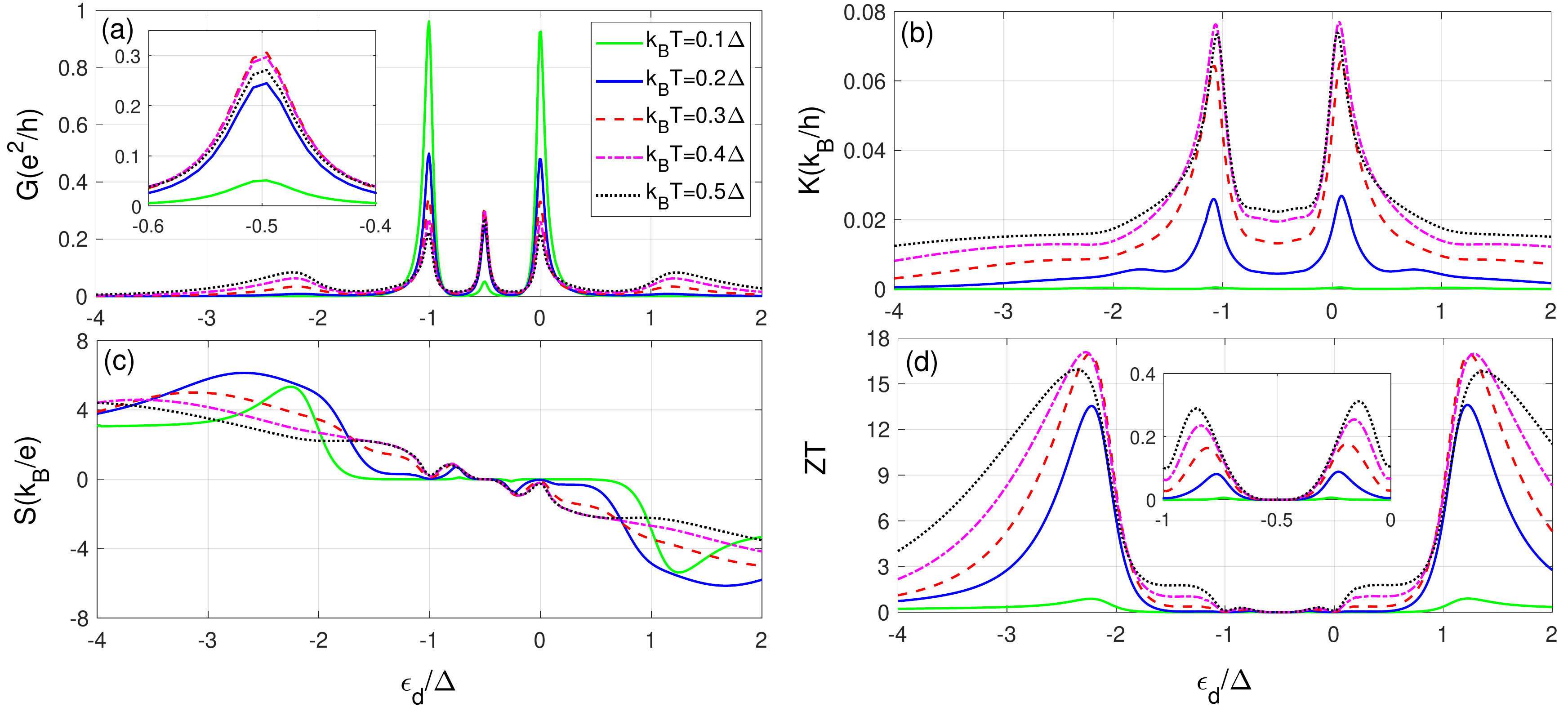}
\caption {Variation of $G$, $K$, $S$, and $ZT$ with the QD energy level $\epsilon_d$ in the linear response regime for several values of the background temperature/thermal energy ($k_BT\geq\Gamma_N$). The other parameters are : $\Gamma_S=\Gamma_N=0.1\Delta$ and $U=\Delta$. Inset in (a) and (d) shows the closeup view of the subgap region.}
\label{fig:4}
\end{figure*}
\begin{figure*}[!htb]
\includegraphics
  [width=0.85\hsize]
  {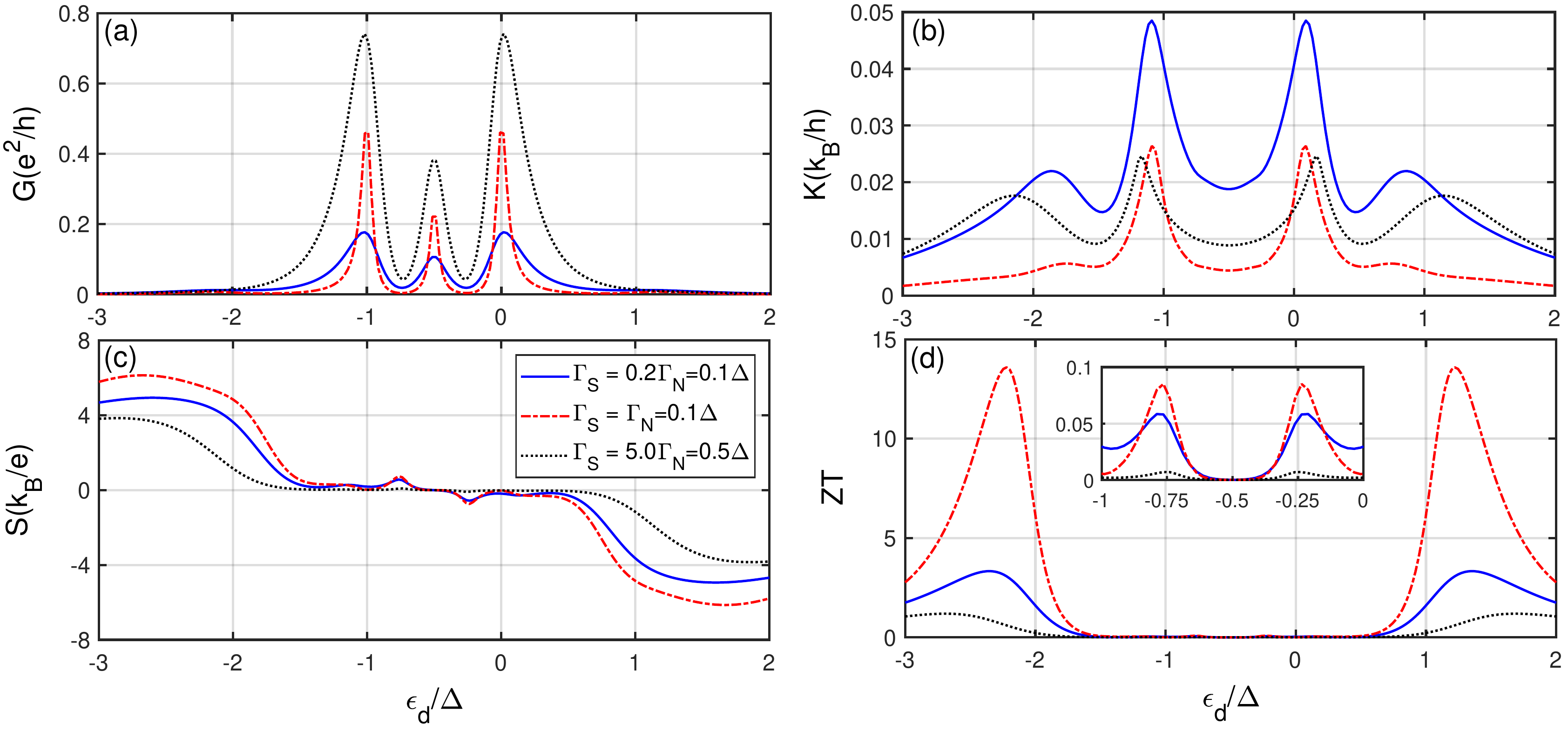}
\caption {Variation of $G$, $K$, $S$, and $ZT$ with the QD energy level $\epsilon_d$ in the linear response regime for three different tunneling coupling ratios ($\Gamma_S/\Gamma_N$). The other parameters are : $k_BT=0.2\Delta$  and $U=\Delta$. Inset in Fig.5(d) shows the closeup view of the subgap region.}
\label{fig:5}
\end{figure*}
By using above linear response relations (\eqref{eq:17}-\eqref{eq:20}) we numerically analyse the thermoelectric transport quantities ($G$,$K$,$S$ and $ZT$) as a function of QD energy level $\epsilon_d$ for several values of parameters $U$, $k_BT$, $\Gamma_S/\Gamma_N$ and $\Delta$.\\
Fig.\ref{fig:2}(a) shows the linear electrical conductance $G$ as a function of the QD energy level $\epsilon_d$ for several values of on-site Coulomb interaction $U$. For non-interacting QD ($U=0$), the electrical conductance $G$ shows a sharp peak centred at Fermi level $\epsilon_d=0$. This peak corresponds to the resonant Andreev tunnelling through the spin-degenerate QD energy level. For interacting QD ($U>0$), two effective levels are lying at $\epsilon_d$ and $\epsilon_d+U$. The electrical conductance now shows three subgap peaks, and each corresponds to the Andreev tunnelling. The side peaks are located at resonance energies $\epsilon_d=0$ and $\epsilon_d=-U$, while the central peak is located at the particle-hole symmetry point ($\epsilon_d=-U/2$). The side peaks corresponding to resonances when either $\epsilon_d=0$ or $\epsilon_d+U=0$ crosses the Fermi level and the height of these peaks are independent of $U$. While the central peak arises from the two-level Andreev tunnelling process i.e. Andreev tunnelling occurs via $\epsilon_d=-U/2$ and $\epsilon_d+U=U/2$ effective levels. The height of this central peak is suppressed as intradot Coulomb repulsion increases because the effective levels move apart from the Fermi energy with increasing $U$ and thus reducing the Andreev tunnelling amplitude. Apart from these subgap peaks, there is a small contribution from the single particle or quasi-particle tunnelling close to the superconducting gap edge, i.e., at $\epsilon_d=\Delta$ and $\epsilon_d=-(\Delta+U)$(inset \ref{fig:2}(a)).\\
In Fig.\ref{fig:2}(c), we plot the corresponding linear thermopower $S$. The curves are asymmetric due to particle-hole symmetry and positive (negative) thermopower shows holes (electrons) as the majority charge carriers. The magnitude of thermopower $|S|$ becomes significant for quasi-particle states near the superconducting gap edge and plays a crucial role in the thermoelectric power generation in the N-QD-S system. Also, notice that the maximum value of thermopower is independent of $U$.  The thermopower becomes zero for $\epsilon_d$ correspond to the Andreev conductance peaks in Fig.\ref{fig:2}(a). This minimization of thermopower occurs because thermal gradient does not give rise to Andreev tunnelling in the linear response regime, and electron and hole current compensate each other at particle-hole symmetry point ($\epsilon_d=-U/2$). However, for $U\geq\Delta$, the additional peaks emerge close to $\epsilon_d=-U/2$ due to quasi-particle tunnelling through one of the two effective levels.\\
Fig.\ref{fig:2}(b) shows the net electronic thermal conductance $K$ as a function QD energy level $\epsilon_d$ for different $U$. The weak $U$ independent peaks near the superconducting gap edge show the quasi-particle thermal conductance, while the in-gap region shows complex $U$ dependence. This behaviour can be better understood by analysing the two terms of Eq.\eqref{eq:19} separately. The inset in Fig.\ref{fig:2}(b) shows  $L_2/T$ (blue dash-dotted), $S^2GT$ (red dotted) and $K$ (solid black) as a function of $\epsilon_d$ for $U=1.5\Delta$. $L_2/T$ represents the thermal conductance between two electrically insulating reservoirs (i.e., without any thermopower) and shows large values for $\epsilon_d$ close to the superconducting gap edge and for $\epsilon_d$ allowing the quasi-particle tunnelling for $U\geq\Delta$. It is also important to note that the quasi-particle tunnelling depend strongly on the background temperature and gives rise to a small thermal conductance for all values of $\epsilon_d$ (non-zero blue dash-dotted curve). $S^2GT$ term represents the thermal conductance corresponding to the thermopower generation, i.e., thermal energy converted into electric energy. Thus low net thermal conductance indicates that considerable thermal energy can be converted into electric energy (i.e., large $ZT$ or thermoelectric efficiency).\\
The figure of merit $ZT$ shows a similar variation with $U$ as shown by thermopower $S$, i.e., $ZT$ become large close to the superconducting gap with a maximum value of $ZT\approx13.6$ [Fig.\ref{fig:2}(d)]. The minima, with vanishing ZT, correspond to the points where $S=0$, i.e., to the Andreev conductance peaks and particle-hole symmetry point. From Eq.\eqref{eq:21}, the linear efficiency corresponding to maximum power output is given by, $\eta_{P_{max}}\approx0.436\;\eta_C$. This efficiency will be useful to compare the linear and non-linear thermoelectric performance of the N-QD-S heat engine.\\
Fig.\ref{fig:3} shows the variation of the figure of merit $ZT$ with the QD energy level $\epsilon_d$ and superconducting energy gap $\Delta$ for different on-site Coulomb interaction $U$. As already discussed previously, $ZT$ for the quasi-particle tunnelling near the superconducting gap edge does not depend on the Coulomb repulsion $U$. However, it is significantly enhanced by the superconducting gap for $1<\Delta<2$. The weak peaks appearing within the gap region due to the finite $U$ effect shows the zero and non-zero $ZT$ regions as a function of $\Delta$. This behaviour arises from the interplay between the quasi-particle tunnelling and Andreev tunnelling processes. For the $\Delta\rightarrow 0$ limit, the results for QD coupled to normal and/or ferromagnetic reservoirs are obtained \cite{Swirkowicz2009,Weymann2013}.\\
Fig.\ref{fig:4} shows the linear thermoelectric quantities as a function of $\epsilon_d$ for several background temperatures ($k_BT\geq\Gamma_N$). In Fig.\ref{fig:4}(a), the quasi-particle contribution to the electrical conductance increases with the background temperature. On the other hand, the side resonant Andreev tunnelling peaks are reduced due to the thermal broadening of the Fermi function in the normal metallic reservoir. Interestingly, the central two-level Andreev tunneling peak is first enhanced and then reduced for $k_BT\geq0.3\Delta$ [see inset \ref{fig:4}(a)]. The origin of such a behaviour may be understood in terms of energy level broadening due to finite temperature effects and reduction of Andreev tunnelling due to the thermal broadening of the Fermi function in the normal metallic reservoir. The temperature dependence of the thermopower $S$ [Fig.\ref{fig:4}(c)] displays a much more complex behaviour and has a minimum for $\epsilon_d$ correspond to the Andreev tunneling peaks. The thermopower corresponds to quasi-particle tunnelling near or outside the superconducting gap is first enhanced and then reduced with the background temperature. The small quasi-particle tunnelling contribution to the thermopower within the gap region becomes constant for $k_BT\geq0.4\Delta$. The thermal conductance $K$ [Fig.\ref{fig:4}(b)] is significantly enhanced with increasing $k_BT$, especially at $\epsilon_d\approx-\Delta$ and $\epsilon_d\approx0$ due to minima in thermopower. The behaviour of $ZT$ is shown in Fig.\ref{fig:4}(d). First, one can see that the thermoelectric efficiency for $k_BT=0.1\Delta$ is relatively small, and it becomes remarkable for $k_BT\geq0.2\Delta$. However, $ZT$ near the superconducting gap edge is slightly reduced for $k_BT=0.5\Delta$. The inset in Fig.\ref{fig:4}(d) shows that the small $ZT$ peaks arising for $U\geq\Delta$ increases with $k_BT$.\\
Fig.\ref{fig:5} shows the linear thermoelectric quantities as a function of $\epsilon_d$ for three tunnelling coupling ratios $\Gamma_S/\Gamma_N$. It is seen in Fig.\ref{fig:5}(a) that the subgap conductance is suppressed for normal reservoir dominate coupling because of the suppression of Andreev tunnelling, while the quasi-particle conductance is enhanced. On the other hand, the net thermal conductance [Fig.\ref{fig:5}(b)] is enhanced for the normal dominate coupling and suppressed for the symmetric couplings ($\Gamma_S=\Gamma_N$). The suppression of $K$ is related to the dominant thermopower $S$ in Fig.\ref{fig:5}(c) for the symmetric coupling. Fig.\ref{fig:5}(d) shows that the combined effect of $G$, $S$, and $K$ causes the significant enhancement of primary peaks in $ZT$ near the superconducting gap edge for symmetric coupling while suppression for superconductor dominate coupling. The inset in Fig.\ref{fig:5}(d) shows that the small $ZT$ peaks arising close to particle-hole symmetry point for $U\geq\Delta$ shows the same coupling dependence as the primary peaks. However, for normal dominate coupling ($\Gamma_S<\Gamma_N$), $ZT$ becomes finite at $\epsilon_d=0$ and $\epsilon_d=-\Delta$ as a result of reduced Andreev tunnelling conductance.
\FloatBarrier
\subsection{Non-Linear regime : Voltage biasing without temperature gradient}
\label{sec:Voltage Driven}
\begin{figure*}[!htb]
\includegraphics
  [width=0.85\hsize]
  {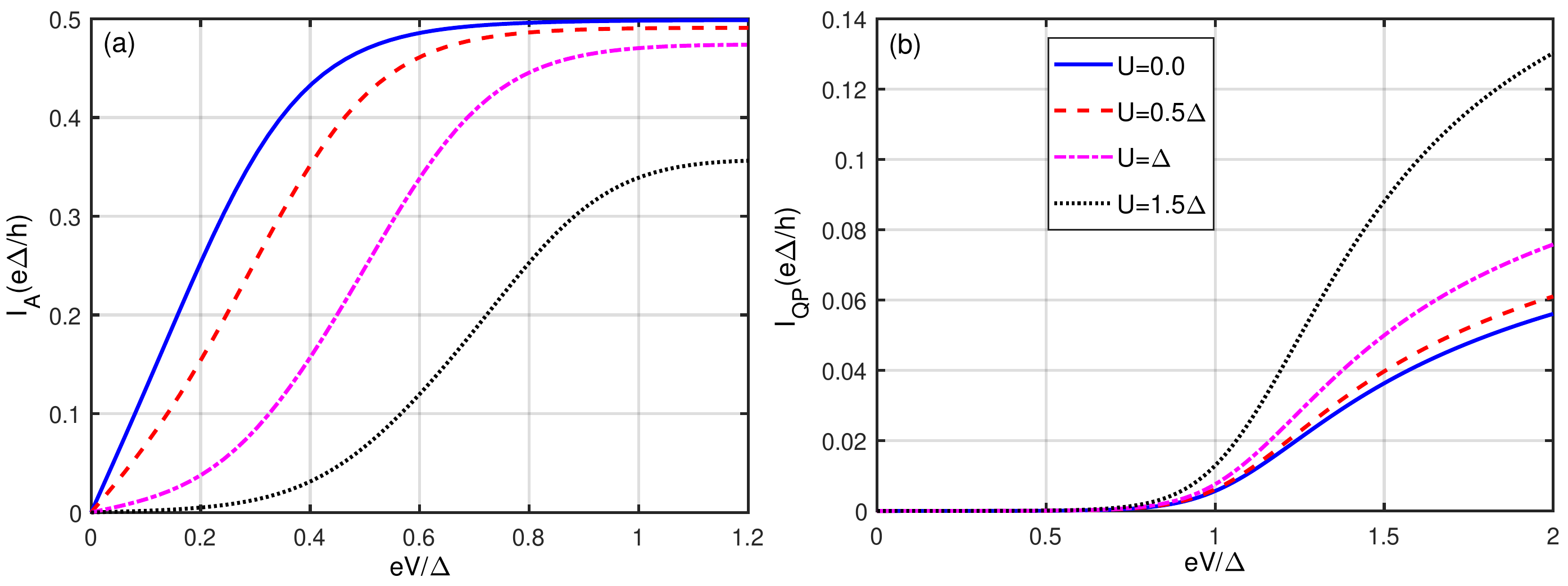}
\caption {(a) Andreev current $I_A$, (b) Quasi-particle current $I_{QP}$ versus voltage for several values of on-site Coulomb interaction $U$ with $\Gamma_N=0.1\Delta$, $\Gamma_S=0.5\Delta$ and $k_BT=0.1\Delta$ at $\epsilon_d=-U/2$.}
\label{fig:6}
\end{figure*}
\begin{figure*}[!htb]
\includegraphics
  [width=0.85\hsize]
  {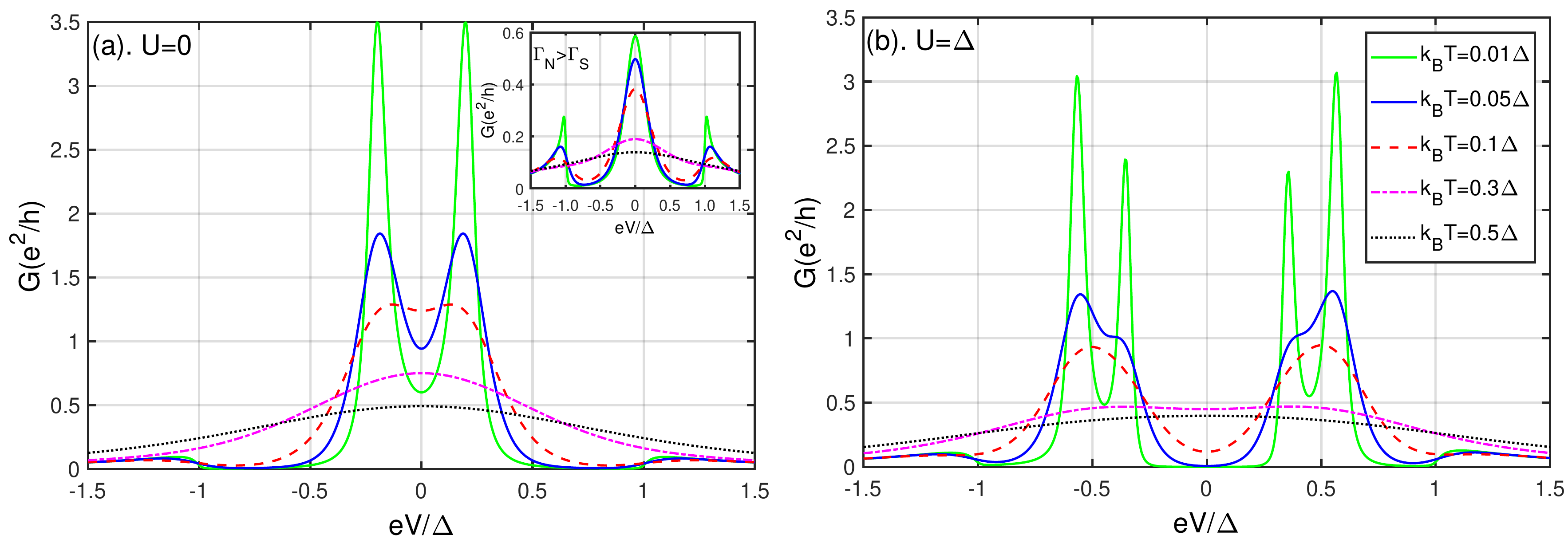}
\caption {Non-linear electrical conductance as a function of bias voltage for different values of the background temperature $k_BT$ with (a) $U=0.0$ and (b) $U=\Delta$. The other parameters are $\Gamma_N=0.1\Delta$, $\Gamma_S = 0.5\Delta$ at $\epsilon_d=-U/2$. The inset shows the electrical conductance as a function of bias voltage for $\Gamma_N=0.5\Delta$ and $\Gamma_S=0.1\Delta$.}
\label{fig:7}
\end{figure*}
 \begin{figure*}[t]
\includegraphics
  [width=0.85\hsize]
  {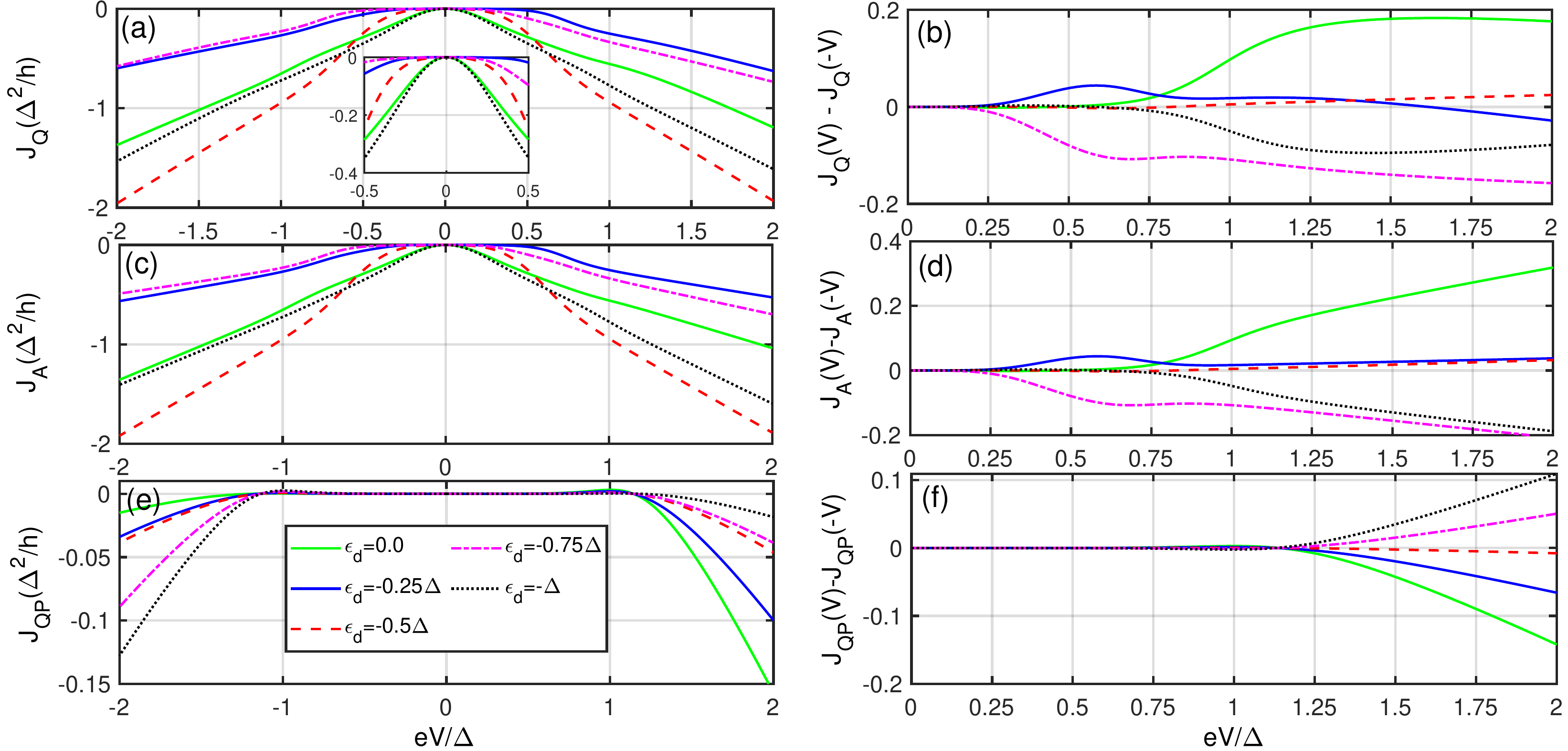}
\caption {{\bf{$\Gamma_S > \Gamma_N$}} case : Fig. (a), (c) and (e) shows the total heat current $J_Q$, Andreev heat current $J_A$ and quasi-particle heat current $J_{QP}$ respectively as a function of applied voltage $eV$ for isothermal reservoirs ($\theta=0$) for several values of the QD energy level position. Fig. (b), (d) and (f) shows the variation of total, Andreev and quasi-particle power rectification with applied biasing. The other parameters are $\Gamma_N=0.1\Delta$, $\Gamma_S = 0.5\Delta$, $k_BT=0.1\Delta$ and $U=\Delta$.}
\label{fig:8}
\end{figure*}
\begin{figure*}[t]
\includegraphics
  [width=0.85\hsize]
  {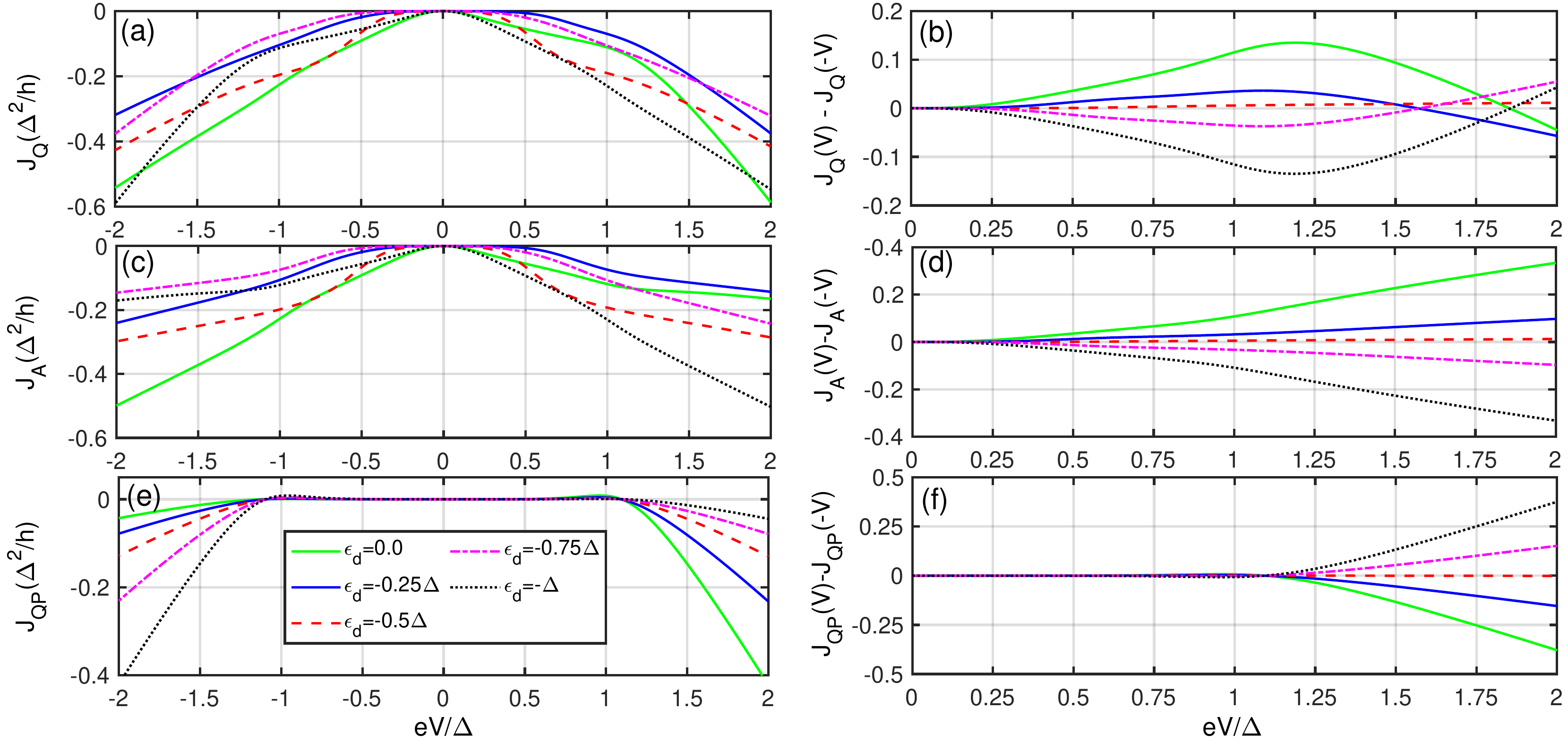}
\caption {{\bf{$\Gamma_S < \Gamma_N$}} case : $\Gamma_N=0.5\Delta$, $\Gamma_S = 0.1\Delta$ and other parameters are same as in Fig.\ref{fig:8}.}
\label{fig:9}
\end{figure*}
\begin{figure*}[!htb]
\includegraphics
  [width=0.85\hsize]
  {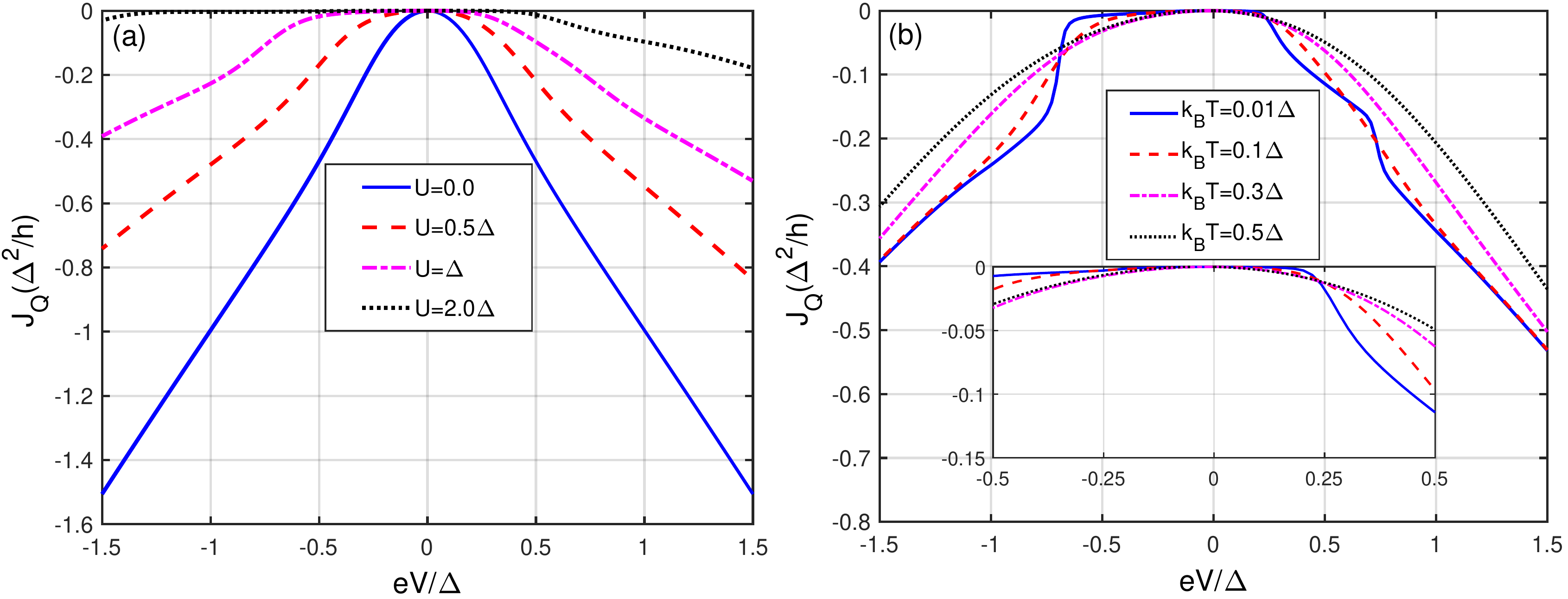}
\caption { Variation of total heat current $J_Q$ as a function of applied voltage $eV$ for several values of (a) Coulomb interaction $U$ with $k_BT=0.1\Delta$ and (b) background thermal energy $k_BT$ with $U=\Delta$. The other parameters are $\Gamma_N=0.1\Delta$, $\Gamma_S = 0.5\Delta$,  and $\epsilon_d=-3U/4$.}
\label{fig:10}
\end{figure*}
In this subsection, we study the non-linear electrical current and electronic contribution to heat current as a function of bias voltage. We also study the asymmetric heat dissipation, previously studied for the N-QD-N system \cite{Lee2013,Sierra2014}. For voltage-driven case with isothermal reservoirs (i.e. $\theta=0$) we set Fermi energy $\mu_f=0$ as the reference point, and consider $\mu_N=eV$ and $\mu_S=0$.\\
We first briefly study the linear electrical current and conductance as a function of voltage biasing for different intradot Coulomb repulsion $U$ and background temperature $k_BT$. In Fig.\ref{fig:6}, we present the current-voltage characteristic curves at the particle-hole symmetry point $\epsilon_d=-U/2$ for several values of intradot Coulomb repulsion $U$. The subgap Andreev current $I_A$ is suppressed with the increasing $U$ and become constant for $eV\geq\Delta$ [Fig.\ref{fig:6}(a)]. In addition, we can see the enhanced non-linear behaviour of Andreev current with increasing $U$. The quasi-particle current $I_{QP}$ for $eV\geq\Delta$ increases with $U$ [Fig.\ref{fig:6}(b)].\\
Fig.\ref{fig:7} present the differential conductance $G=dI_C/dV$ as a function of biased voltage for different background temperature $k_BT$. For the superconducting dominate coupling $\Gamma_S>\Gamma_N$ [Fig.\ref{fig:7}(a)], the subgap Andreev peaks dominate the transport while the quasi-particle tunnelling peaks near the superconducting gap edge ($|eV|\approx\Delta$) are suppressed. Also, the height of the Andreev conductance peaks are strongly suppressed by thermal fluctuation. Eventually, for $k_BT\geq0.3\Delta$, two peaks are no longer resolved and appear as a single broad peak structure. For normal dominate coupling $\Gamma_S<\Gamma_N$, the subgap Andreev conductance exhibits a zero-bias peak arising because the broadening due to interaction with the normal reservoir exceed the splitting of Andreev peaks [see inset \ref{fig:7}(a)]. Also, note that these subgap Andreev tunnelling peaks are now suppressed, while the single quasi-particle tunnelling peaks are slightly enhanced relative to $\Gamma_S>\Gamma_N$ case. For relatively low temperatures, the competition between Andreev tunnelling and finite Coulomb interaction may lead to the additional splitting of the $U=0$ subgap Andreev peaks and develop a local minimum close to zero biasing [see Fig.\ref{fig:7}(b)]. The two outer Andreev conductance peaks may not be visible in the spectroscopy experiments since they can merge with the outer quasi-particle continuum with increasing $U$ \cite{Kumar2014}. As the temperature increases, these peaks convert into two broad peaks, and eventually, the Coulomb interaction effect becomes negligible for $k_BT\geq0.3\Delta$. At low enough temperature ($T<T_K$, where $T_K$ is the Kondo temperature), a prominent zero-bias Kondo peak may develop at the local minima, which is out of the scope of the present analysis of the Coulomb blockade regime.\\
In addition to the charge current electrons also carry energy. Thus voltage biasing also leads to the Peltier effect and Joule heating effect. The former describes a reversible heat flow for low voltages in isothermal condition. In other words, the electrical current flowing through the QD connecting source and drain reservoirs will emit or absorb heat per unit time to balance the difference in the chemical potential of the two reservoirs. Also, in the Peltier effect, the heating or cooling of the system depends on the bias polarity or direction of current flow. On the other hand, in the Joule heating effect, the charge transport is always accompanied by irreversible heat dissipation. There is no subgap Peltier effect in the linear response regime due to vanishing Andreev heat current ($J_A\rightarrow0$) because electron and hole heat energy cancel each other. Beyond the linear response regime, the subgap Andreev heat current can become finite and play an vital role in asymmetric heat dissipation and rectification for low bias voltages.\\
Fig.\ref{fig:8} and Fig.\ref{fig:9} shows the total heat current ($J_Q$), Andreev heat current ($J_A$), quasi-particle heat current ($J_{QP}$), and corresponding asymmetric heat dissipation for different values of $\epsilon_d$ in the superconducting dominate coupling ($\Gamma_S>\Gamma_N$) and normal dominate coupling ($\Gamma_S<\Gamma_N$) respectively. The Andreev heat current is large for superconducting dominate coupling $\Gamma_S>\Gamma_N$ than the normal dominate coupling $\Gamma_S<\Gamma_N$. On the other hand, the quasi-particle heat current (for $|eV|\geq\Delta$) is enhanced in the normal dominate coupling. For $|eV|<\Delta$ the quasi-particle contribution to heat current ($J_{QP}$) is almost zero for low background thermal energy. For low voltage biasing and specific QD energy level the total heat current can be rectified and act as a thermal diode [inset in Fig.\ref{fig:8}(a) for $\epsilon_d=-0.75\Delta$ and $\epsilon_d=-0.25\Delta$]. However, the Andreev Joule heating effect quickly dominates over the Peltier effect, and eventually, the heat rectification ceases. We also observe that, $J_Q$, $J_A$, and $J_{QP}$ are symmetric around $eV=0$ for the particle-hole symmetry point similar to the N-QD-N system \cite{Sierra2014}. However for superconducting dominate coupling $\Gamma_S>\Gamma_N$ the invariance of heat currents $J_Q$, $J_A$, and $J_{QP}$ under the simultaneous transformation of $eV\rightarrow-eV$ and $\epsilon_d\rightarrow-\epsilon_d-U$ is no longer valid [see Fig.\ref{fig:8}(a),\ref{fig:8}(c) \& \ref{fig:8}(e)]. The invariance is restored for normal dominate coupling $\Gamma_S<\Gamma_N$ [see Fig.\ref{fig:9}(a), \ref{fig:9}(c) \& \ref{fig:9}(e)].\\
In Fig.\ref{fig:8}(b), \ref{fig:8}(d) and \ref{fig:8}(f) (Fig.\ref{fig:9}(b), \ref{fig:9}(d) and \ref{fig:9}(f)) we have shown the respective asymmetric heat dissipation for different values of the QD energy level $\epsilon_d$ for $\Gamma_S>\Gamma_N$($\Gamma_S<\Gamma_N$). If the transport is particle-hole symmetric ($\epsilon_d=-U/2$) or $eV\lesssim 0.2\Delta$, then heat is almost equally dissipated for both positive and negative voltage biasing. Thus, in order to have a heating asymmetry, i.e., $J_{Q}(V)-J_{Q}(-V)\neq0$, one needs a certain degree of particle-hole asymmetry (by tuning the QD energy level $\epsilon_d$) and $eV>0.2\Delta$. If the QD energy level $\epsilon_d$ lie above the particle-hole symmetry point then asymmetric heat dissipation correspond to Andreev heat current is always positive, i.e., the dissipation is larger for $eV >0$ than for $eV<0$ and shows complex variation with bias voltage. On the other hand, if the QD energy level $\epsilon_d$ lies below the particle-hole symmetry point, then asymmetric heat dissipation corresponds to Andreev heat current is always negative, i.e., more dissipation for negative voltages as compared to positive voltages. The situation is completely opposite for the asymmetric dissipation corresponds to the quasi-particle heat current. Thus the total asymmetric dissipation changes the sign, i.e., cross the x-axis or voltage-axis only when quasi-particle contribution becomes finite (i.e., $eV>\Delta$). For normal dominant coupling ($\Gamma_S<\Gamma_N$), the asymmetric heat dissipation is perfectly symmetric about the x-axis, i.e., particle-hole symmetry point, and cross x-axis in a more controllable manner for different values of $\epsilon_d$. The study of the asymmetric heat dissipation can be helpful in the design of superconductor-QD-based nano-devices with controllable dissipation.\\
Fig.\ref{fig:10}(a) shows the total heat current versus biasing voltage for different Coulomb interaction $U$. The heat current is large for non-interacting QD ($U=0$) with equal dissipation in both transport directions due to particle-hole symmetry. When $U$ becomes finite, the heat current is reduced and strongly depend upon the transport direction. Thus asymmetric heat dissipation arises due to the Coulomb blockade effect away from the particle-hole symmetry. Also, heat rectification can be possible at relatively larger voltage biasing for strong Coulomb interaction ($U=2\Delta$). The origin of the asymmetries in the heat current with respect to the bias polarity can be explained as follows: For $|eV|<\Delta$, Andreev heat current dominates the heat transport beyond the linear response. However, at the particle-hole symmetric point ($\epsilon_d=-U/2$), the electron and hole energies are located symmetrically around the Fermi level $\epsilon_f$, which results in symmetric Andreev heat current for positive and negative bias. Tuning the QD energy level away from the particle-hole symmetry point causes one of the effective levels to shift close to the Fermi energy ($\epsilon_f$) while the other moves further away. Thus heat dissipation can be larger or smaller for a given bias as compared to its reverse value, depending on the position of QD level with respect to Fermi level{\cite{zotti2014}}. In Fig.{\ref{fig:10}}, two effective QD levels are located at $\epsilon_d=-3U/4$ and $\epsilon_d+U=U/4$. Thus Andreev tunnelling amplitude is higher (large heat current) in the upper part of the transport channel (effective level close to $\epsilon_f$) for positive biasing, and more heat is dissipated for negative biasing. Also, increasing the Coulomb interaction $U$ the Andreev heat current is reduced because the effective QD energy levels moves further away from $\epsilon_f$. Fig.\ref{fig:10}(b) shows the effect of thermal fluctuation on the total heat current for finite $U$. The heat current first increases with $k_BT$ for low voltage biasing, but the Joule heating effect quickly dominates, and as a result, the heat current starts to decrease with increasing $k_BT$. The inset in Fig.\ref{fig:10}(b) shows that the heat rectification for low biasing become more effective at the low background temperature or thermal fluctuation ($J_Q(V=0.5\Delta)\approx10J_Q(V=-0.5\Delta)$ for $k_BT=0.01\Delta$ and $J_Q(V=0.5\Delta)\approx5J_Q(V=-0.5\Delta)$ for $k_BT=0.1\Delta$). It is also interesting to explore the combined influence of electric and thermal fields on heat transport in N-QD-S system and will be address in future work.
\subsection{Non-Linear regime : Thermoelectric heat engine}
\label{sec:Heat Engine}
\begin{figure}[!htb]
\includegraphics
  [width=0.85\hsize]
  {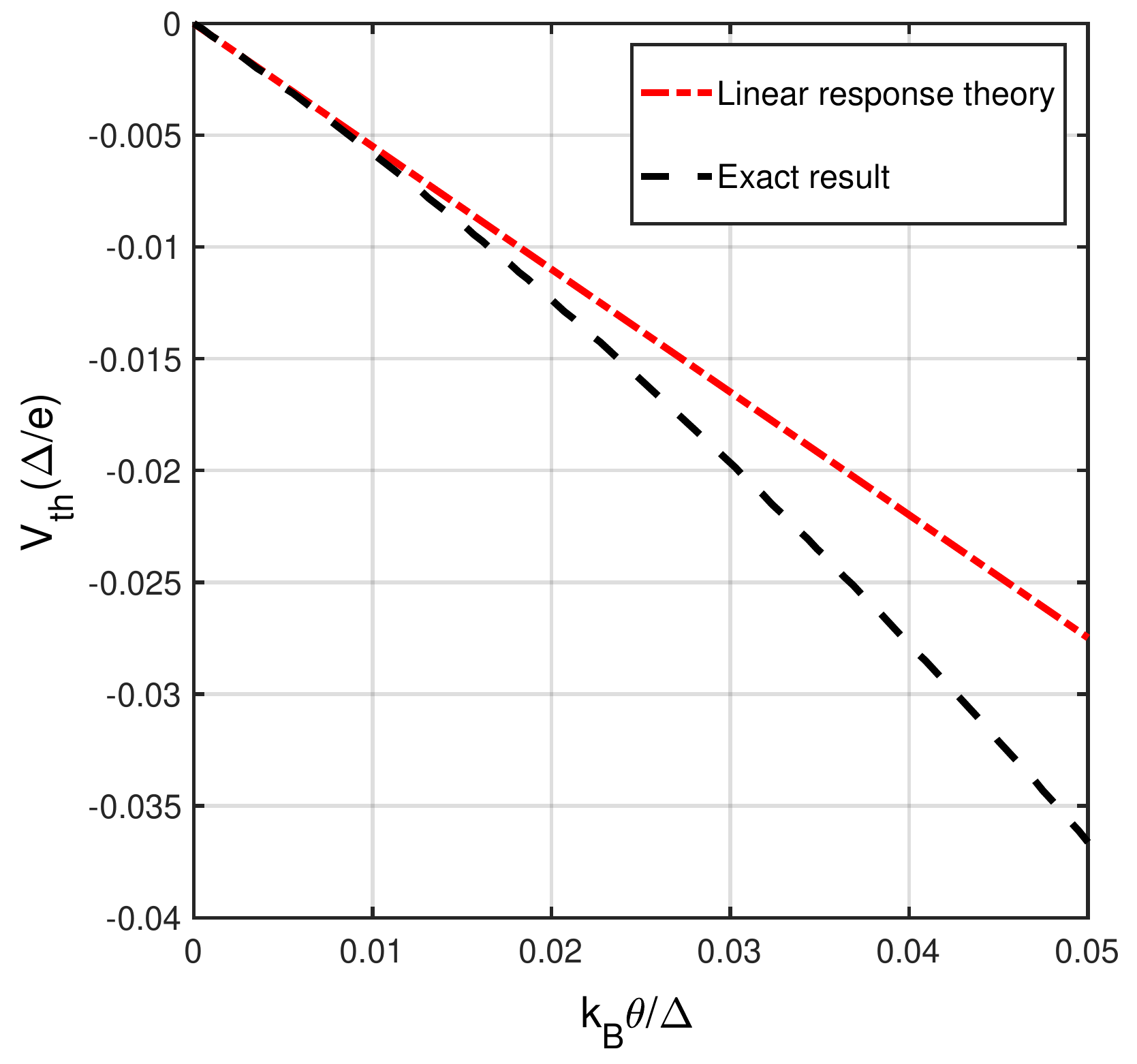}
\caption {Thermovoltage ($V_{th}$) vs thermal gradient $\theta$ calculated by using linear response theory (Eq.\eqref{eq:18}) and exact equation (Eq.\eqref{eq:28}) for a correlated QD ($U=\Delta$) with $\epsilon_d = 0.5\Delta$, $\Gamma_N=\Gamma_S=0.1\Delta$ and $k_BT=0.2\Delta$.}
\label{fig:11}
\end{figure}
\begin{figure*}[!htb]
\includegraphics
  [width=0.85\hsize]
  {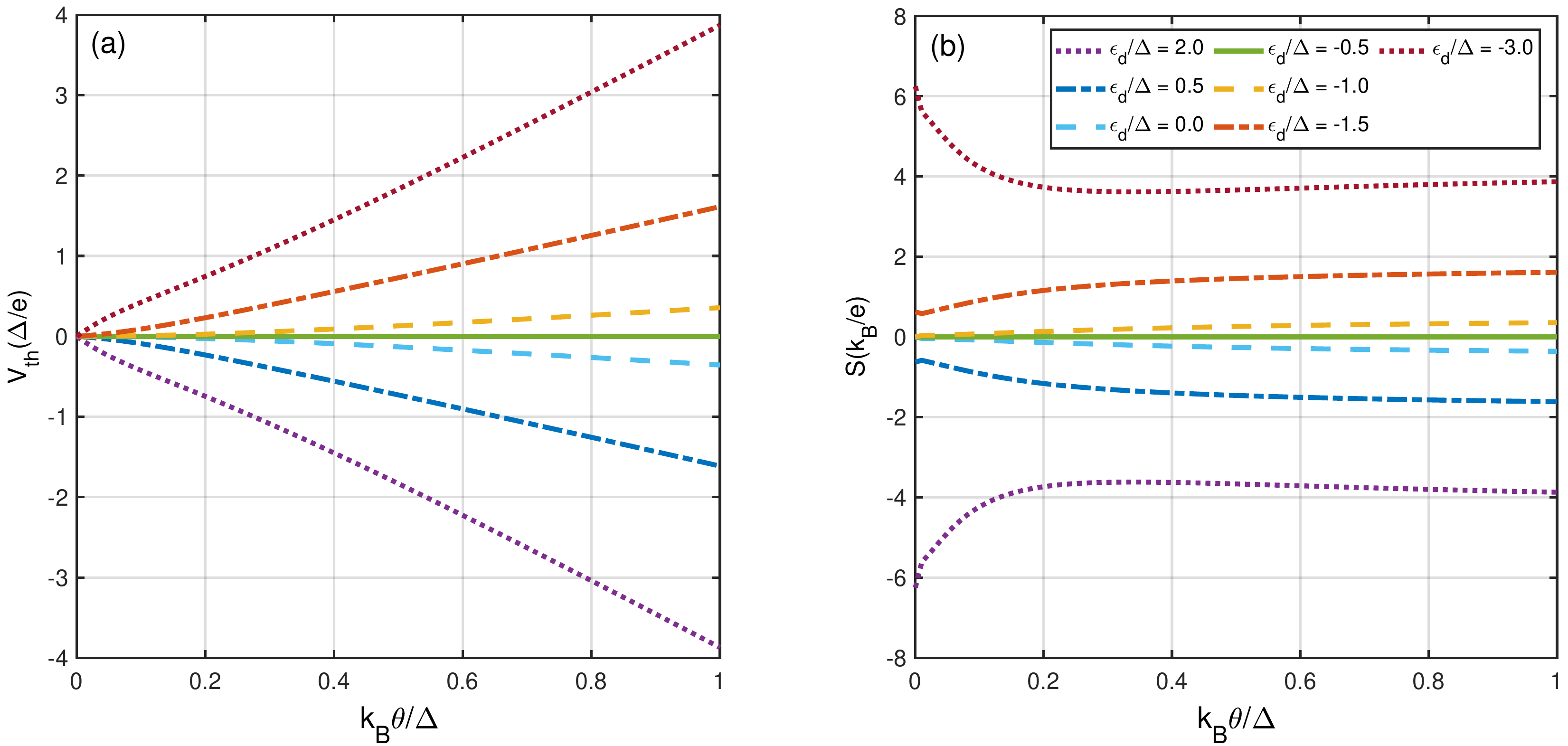}
\caption {(a) Thermovoltage ($V_{th}$) and (b) thermopower ($S$) as a function of thermal gradient ($\theta$) for several values of dot energy level ($\epsilon_d$). The other parameters are $U=\Delta$, $\Gamma_N=\Gamma_S=0.1\Delta$ and $k_BT=0.2\Delta$.}
\label{fig:12}
\end{figure*}
\begin{figure}[!htb]
\includegraphics[scale=0.45]
  {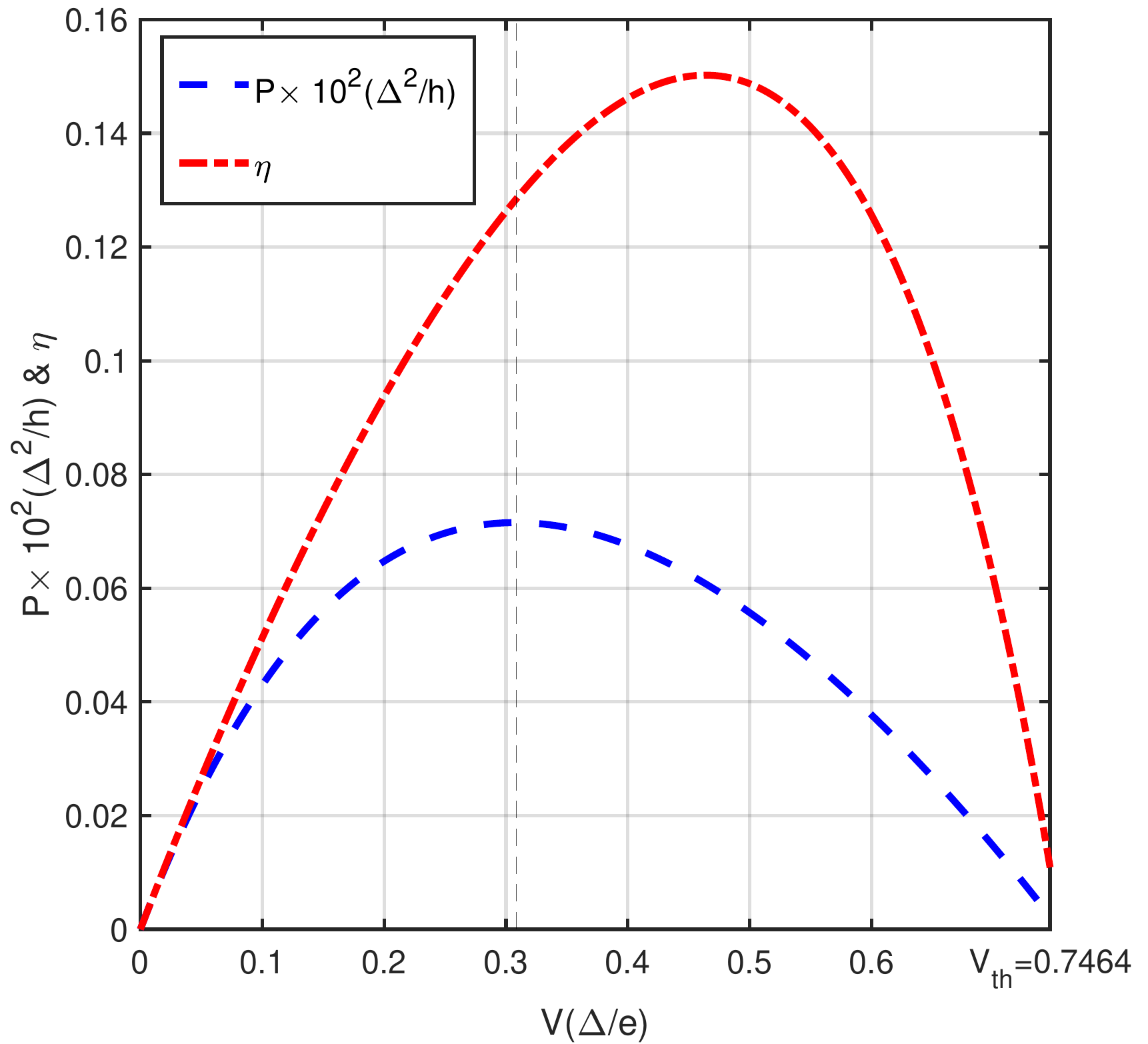}
\caption {Variation of power output $P$ and corresponding efficiency $\eta$ as a function of applied bias voltage. Black dashed line parallel to y-axis shows the maximum power output and corresponding efficiency. The parameters are : $U=\Delta$, $\epsilon_d=-U$, $k_B\theta=0.2\Delta$, $k_BT=0.2\Delta$ and $\Gamma_N=\Gamma_S=0.1\Delta$.}
\label{fig:13}
\end{figure}
\begin{figure*}[!htb]
\includegraphics
  [width=0.85\hsize]
  {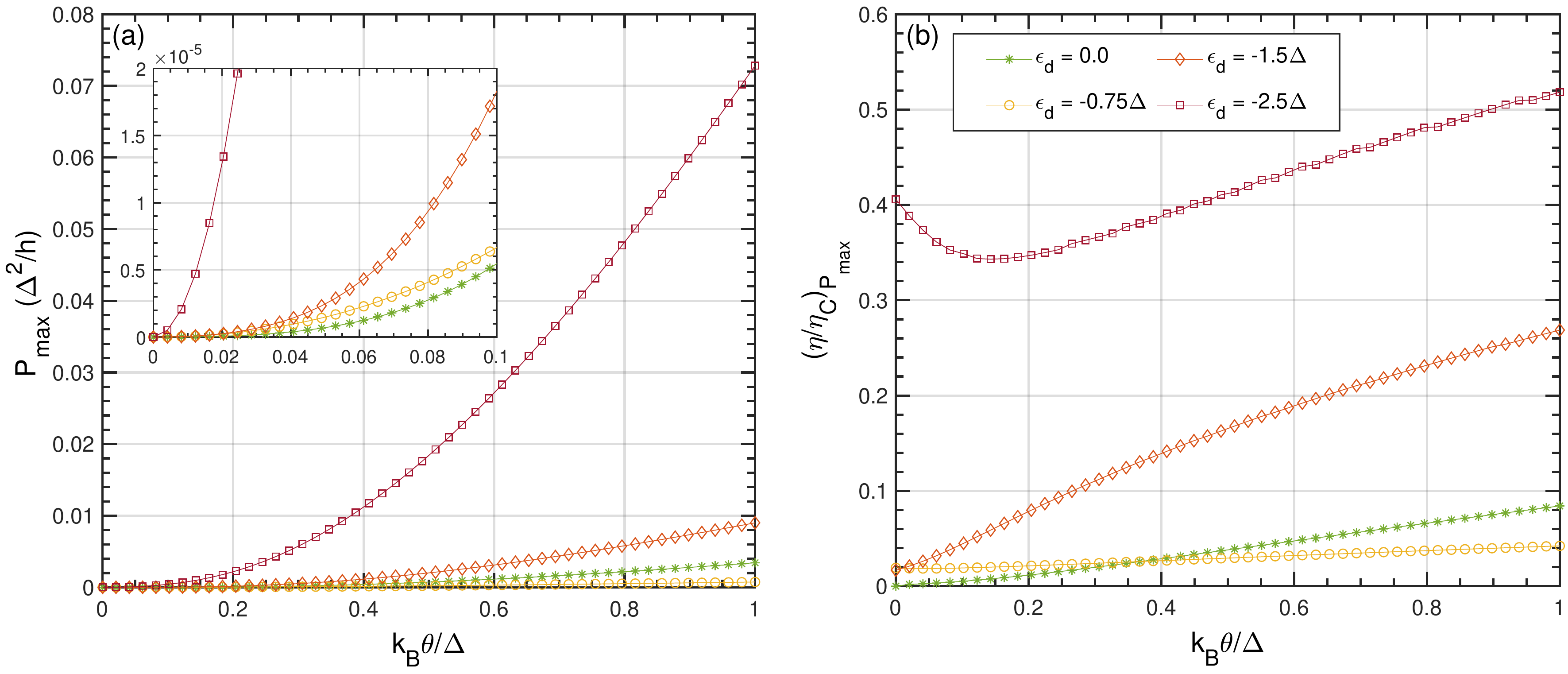}
\caption {(a) Variation of maximum output power $P_{max}$ with the temperature gradients $k_B\theta$ for four different QD energy levels. The inset shows the close-up view for small $\theta$. (b) Variation of  relative efficiency corresponding to maximum power output with the temperature gradients $k_B\theta$ for four different QD energy levels. The other parameters are : $\Gamma_S=\Gamma_N=0.1\Delta$, $k_BT=0.2\Delta$ and $U = \Delta$.}
\label{fig:14}
\end{figure*}
In order to use N-QD-S system as a heat engine or power generator at finite thermal gradient, we consider $T_N=T+\theta$ and $T_S=T$. This temperature difference ($\theta>0$) between normal and superconducting reservoirs generate a finite thermovoltage $V_{th}=(\mu_N-\mu_S)/e$. It is clear from Fig.\ref{fig:11} that, the linear response thermovoltage calculated by using Eq.\eqref{eq:18} quickly deviates from exact thermovoltage obtained from the numerical solution of Eq.\eqref{eq:28}. Thus linear response theory results from section \hyperref[sec:Linear]{\Romannum{3}.A} are only valid for very low thermal gradient energies $k_B\theta\lesssim 0.01\Delta$ and it is not reliable to judge the thermoelectric properties for relatively larger $k_B\theta$ from the linear response regime.\\
In Fig.\ref{fig:12}(a), we plot the non-linear thermovoltage $V_{th}$ as a function of thermal gradient ($\theta$) for several values of $\epsilon_d$. The corresponding thermopower $S$ is shown in Fig.\ref{fig:12}(b). Both thermovoltage and thermopower are zero for all $\theta$ at particle-hole symmetric point ($\epsilon_d=-U/2$). By tuning the QD energy level above or below the particle-hole symmetry point, $|V_{th}|$ become a monotonic function of $\theta$ for all values of $\epsilon_d$. This behaviour of $V_{th}$ as a function of $\theta$ is consistent with previous result \cite{Hwang2015} and is different from N-QD-N system in which $V_{th}$ can become zero and changes sign for certain $\epsilon_d$ at nonzero $\theta$ \cite{Svensson2013,Sierra2014}. The thermopower $|S|$ for low thermal gradient is large near the superconducting gap due to quasi-particle tunnelling, which decreases with $\theta$ for $k_B\theta<0.2\Delta$, before attaining a constant value for large $k_B\theta$. If $\epsilon_d$ lies within the superconducting energy gap (i.e., $-(\Delta+U)<\epsilon_d<\Delta$), then thermopower $|S|$ is relatively small due to suppression of quasi-particle tunnelling for low $k_BT$. Also, $|V_{th}|$ and $|S|$ for $\epsilon_d = 0$ and  $\epsilon_d = -\Delta$ are vanishingly small for low thermal gradient ($k_B\theta\lesssim 0.4\Delta$). Andreev tunnelling itself does not exhibit the Seebeck effect (because no thermopower is generated) but it suppress $|V_{th}|$ and $|S|$ indirectly due open circuit condition.\\
The electrical power given by Eq.\eqref{eq:29} is zero when no bias voltage is applied (i.e., zero load resistance for $V=0$ ) or when $V=V_{th}$ (i.e., bias at which $I_C=0$ or infinite load resistance). Thus maximal power output is at a bias voltage between $V=0$  and $V=V_{th}$ [for example, see Fig.\ref{fig:13}]. The efficiency corresponding to the maximum power output as a function of applied bias $V$ is also shown in Fig.\ref{fig:13}. Furthermore, it can be seen that the efficiency at the maximum power output is smaller than the maximal efficiency that can be achieved by tuning the applied bias. Next, we consider how this maximum value of power output and corresponding efficiency changes when the thermal gradient $\theta$ is varied. Fig.\ref{fig:14}(a) shows that, the maximum power output $P_{max}$ can be amplified by tuning the QD energy level ($\epsilon_d$) above and below the particle-hole symmetry point. $P_{max}$ increases by a large factor as $\epsilon_d$ approaches the quasi-particle states near the superconducting energy gap edge. We also observe the increase of  $P_{max}$ as a function of $\theta$. This increase is more dramatic for the quasi-particle tunnelling peaks near the superconducting gap edge. For example the maximum output power at $\epsilon_d=-2.5\Delta$ is of the order of $pW$ and increased approximately by a factor of $45$ from $k_B\theta\approx0.2\Delta$ to $k_B\theta\approx\Delta$. Furthermore, for $k_B\theta<0.1\Delta$ the order of maximum power is $fW$ if $\epsilon_d$ lies within the superconducting gap region. For large $k_B\theta$, we observe $P_{max}$ of the order $0.1pW$ at $\epsilon_d=0$ (i.e. correspond to the Andreev tunnelling peaks). Fig.\ref{fig:14}(b) shows the variation of efficiency at the maximum power output with thermal gradient $\theta$. It is seen that there is a significant improvement in the $\eta_{P_{max}}$ for thermally induce quasi-particle tunnelling peaks within the sub-gap region. While $\eta_{P_{max}}$ for the quasi-particle tunnelling close to superconducting gap edge shows a minima for $k_B\theta\approx0.2\Delta$ and then slightly increases from $\eta_{P_{max}}\approx0.41\eta_C$ to $\eta_{P_{max}}\approx0.52\eta_C$ for $k_B\theta>0.5\Delta$.
\subsection{Influence of proximity induced gap on the transport through QD}
\label{sec:induced superconducting gap}
\begin{figure*}[!htb]
\includegraphics
  [width=0.85\hsize]
  {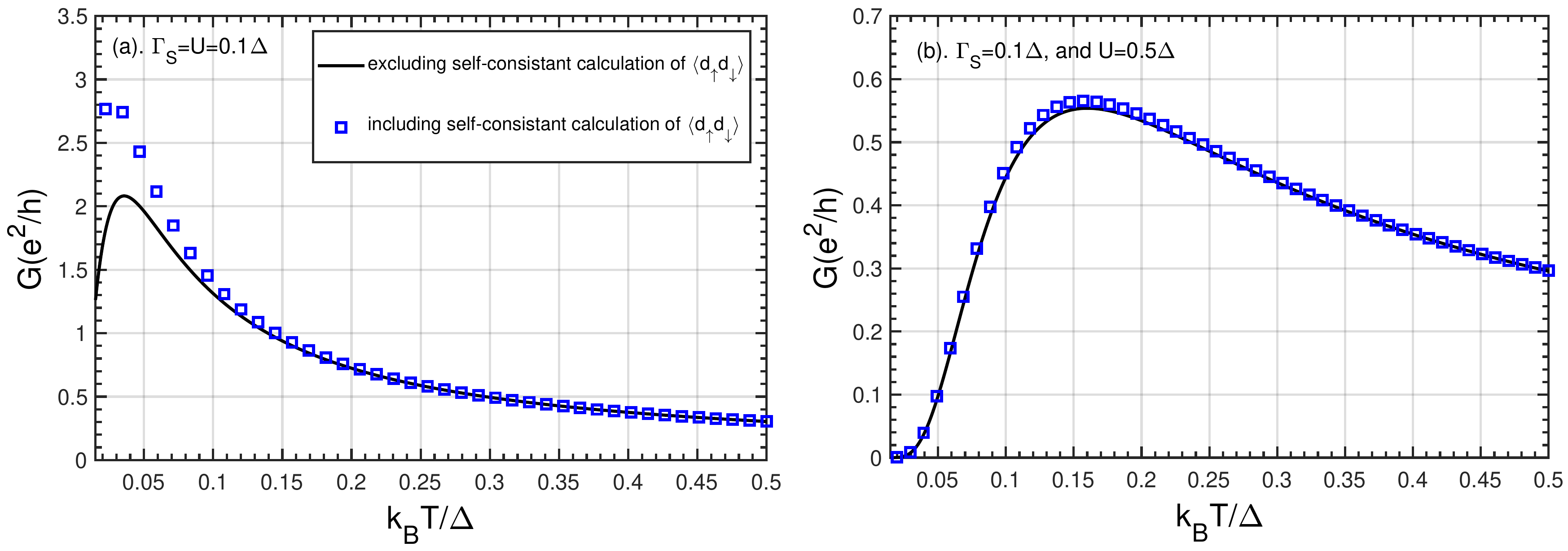}
\caption {Variation of linear electrical conductance $G$ with the background thermal energy $k_BT$ with (blue square) and without (solid black line) the self-consistent calculation of $\langle{d_{\uparrow}d_{\downarrow}}\rangle$ for (a) $\Gamma_S=U$ and  (b) $\Gamma_S=0.2U$. The other parameters are : $\Gamma_N=\Gamma_S=0.1\Delta$, and $\epsilon_d = -U/2$.}
\label{fig:15}
\end{figure*}
The expectation value $\langle{d_{\uparrow}d_{\downarrow}}\rangle$ is the measure of the robustness of the superconducting proximity effect at the QD site and quantifies the induced on-site superconducting energy gap $\Delta_d$. The earlier studies of thermoelectric transport through hybrid superconductor-QD nano-structures ignores the self-consistent evaluation of $\langle{d_{\uparrow}d_{\downarrow}}\rangle$\cite{Hwang2016b,Hwang2017,Barnas2017}.\\
In this subsection we evaluate the expectation value $\langle{d_{\uparrow}d_{\downarrow}}\rangle$ self-consistently along with occupancy $\langle{n_{\sigma}}\rangle$ and justify our assumption about the exclusion of $\langle{d_{\uparrow}d_{\downarrow}}\rangle$. 
The self-consistent equation for $\langle{d_{\uparrow}d_{\downarrow}}\rangle$ is given by 
\begin{equation}
\langle{d_{\uparrow}d_{\downarrow}}\rangle=\frac{-i}{2\pi}\int^{\infty}_{-\infty}G^{<}_{d,12}(\omega) d\omega
\end{equation}
where $G^{<}_{d,12}$ is the off-diagonal element of the lesser Green's function (Eq.\eqref{eq:15}).\\
In Fig.\ref{fig:15} we have shown the effect of proximity induced superconducting gap on the linear electrical conductance $G$ as a function of background thermal energy ($k_BT\geq0.015\Delta$) for two different values of $\Gamma_S/U$. It is clear that self-consistent evaluation of $\langle{d_{\uparrow}d_{\downarrow}}\rangle$ become significant only at low thermal energies $k_BT\leq 0.05\Delta$ with $\Gamma_S\approx U$ (Fig.\ref{fig:15}.(a)). On the other hand for $\Gamma_S<U$ (Fig.\ref{fig:15}.(b)) the effect of  $\langle{d_{\uparrow}d_{\downarrow}}\rangle$ is negligible because strong Coulomb blockade effect on the dot prevent the double occupancy and eventually suppressing the effect of $\langle{d_{\uparrow}d_{\downarrow}}\rangle$ on the linear electrical conductance. Similarly one can also check that excluding the self-consistent equation for proximity induced superconducting gap doesn't change other electric and thermoelectric transport quantities in linear and non-linear regime for the parameter regimes considered in the present work.
\section{Conclusion and outlook}
\label{sec:Conclusion}
We have discussed the non-equilibrium steady-state thermoelectric transport properties of an elementary single-level QD coupled to normal metallic and BCS superconducting reservoirs in the presence of intra-dot Coulomb correlation. In the linear response regime, we studied the Coulomb interaction, background temperature, superconducting energy gap, and dot-reservoir coupling dependence of the thermoelectric quantities as a function of quantum dot energy level. We found that the magnitude of quasi-particle tunnelling near the superconducting gap edge is $U$-independent and dominates the thermoelectric transport properties. On the other hand, subgap Andreev tunnelling only shows large electrical conductance and no contribution to the thermopower and heat current in the linear response regime. However, Andreev tunnelling plays a significant role in the suppression of thermopower within the sub-gap region. In the non-linear response regime, the subgap Andreev heat current generated in response to a voltage bias becomes significant. It plays an essential role in asymmetric heat dissipation and low-bias thermal rectification. We found that the asymmetric heat dissipation can become zero and changes sign only when quasi-particle heat current become finite, i.e., $eV>\Delta$. This behaviour can be useful for the design of nano-devices with controlled heat dissipation. Further, the N-QD-S system can act as a thermal diode at low voltage biasing with enhanced rectification for strong Coulomb interaction and low background temperature. Finally, the thermovoltage, thermopower, maximum power output, and corresponding efficiency for the N-QD-S device was investigated in the non-linear transport regime. The efficiency corresponding to the maximum power output reaches the value $\eta_{P_{max}}\approx0.5\eta_C$ close to the superconducting gap edge for a relatively large thermal gradient. We also observe a small power output with corresponding efficiency $\eta_{P_{max}}\lesssim0.08\eta_C$ for moderate and large $k_B\theta$ at $\epsilon_d=0$ as a manifestation of the reduced Andreev tunnelling process. Interestingly, to understand electric and thermoelectric transport properties at finite temperature in the Coulomb blockade regime, the proximity-induced superconducting gap on the QD state does not play any significant role.\\
Thus, the aspects that are of particular interest in the present analysis are the investigation of heat transport in response to voltage bias and the study of the thermoelectric heat engine beyond the linear regime, which has not been received much attention in earlier studies. We believe that the presented analytical and numerical analysis provides the basis for the further in-depth understanding of the non-equilibrium charge and heat transport in hybrid superconductor-QD nanostructures. The related research work may also be taken out addressing linear and non-linear thermoelectric transport through multi-level QD\cite{Zianni2008}, multi-terminal configuration\cite{Michalek2016}, multi-dot configuration\cite{Xu2016,Nie2016}, and the influence of Kondo interaction on the thermoelectric transport through these hybrid nanostructures\cite{Eckern12020}.\\
\begin{acknowledgements}
Sachin Verma, is presently a research scholar at the department of physics IIT Roorkee and would like to acknowledge the financial support from the Ministry of Human Resource Development (MHRD), India, in the form of Ph.D. fellowship.
\end{acknowledgements}

\end{document}